%% file: quijote_ame_pol_3r.tex
\documentclass{aa}  

\usepackage{graphicx}
\usepackage{txfonts}
\usepackage{xcolor}
\usepackage{siunitx}
\usepackage[colorlinks=true,linkcolor=blue,citecolor=blue,filecolor=blue,urlcolor=blue]{hyperref}
{}{\RequirePackage[colorlinks=true,
    linkcolor=blue, citecolor=blue, filecolor=blue, urlcolor=blue]{hyperref}}%

\begin{document}
\nolinenumbers

\title{QUIJOTE scientific results -- XVIII. New constraints on the polarization of the Anomalous Microwave Emission in bright Galactic regions: $\rho$\,Ophiuchi, Perseus and W43}
\titlerunning{QUIJOTE scientific results XVIII}
\subtitle{}

\author
{ R. Gonz\'{a}lez-Gonz\'{a}lez,\inst{1,2}\thanks{E-mail: raul.gonzalez@iac.es (RGG)}
\and
R.~T.~G\'{e}nova-Santos\inst{1,2}\thanks{E-mail: rgs@iac.es (RTGS)}
\and
J.~A.~Rubi\~{n}o-Mart\'{\i}n\inst{1,2}
\and
M.~W.~Peel\inst{3,1,2}
\and
F.~Guidi\inst{4}
\and
C.~H.~L\'{o}pez-Caraballo\inst{1,2}
\and
M.~Fern\'{a}ndez-Torreiro\inst{1,2,5}
\and
R.~Rebolo\inst{1,2,6}
\and
C.~Hern\'andez-Monteagudo\inst{1,2}
\and
D.~Adak\inst{1,2}
\and
E.~Artal\inst{7}
\and
M.~Ashdown\inst{8,9}
\and
R.~B.~Barreiro\inst{10}
\and
F.~J.~Casas\inst{10}
\and
R.~Cepeda-Arroita\inst{1,2}
\and
E. de la Hoz\inst{10,11,12}
\and
A.~Fasano\inst{1,2}
\and
D.~Herranz\inst{10}
\and
R.~J.~Hoyland\inst{1,2}
\and
E.~Mart\'{i}nez-Gonzalez\inst{10}
\and
G.~Pascual-Cisneros\inst{10}
\and 
L.~Piccirillo\inst{13}
\and
F.~Poidevin\inst{1,2}
\and
B.~Ruiz-Granados\inst{14,1,2}
\and
D.~Tramonte\inst{15,1,2}
\and
F.~Vansyngel\inst{1,2}
\and
P.~Vielva\inst{10}
\and
R.~A.~Watson\inst{13}}

\institute{Instituto de Astrof\'{\i}sica de Canarias, E-38200 La Laguna, Tenerife, Spain
\and
Departamento de Astrof\'{\i}sica, Universidad de La Laguna, E-38206 La Laguna, Tenerife, Spain
\and
Imperial College London, Blackett Lab, Prince Consort Road, London SW7 2AZ, UK
\and
Institut d'Astrophysique de Paris, UMR 7095, CNRS \& Sorbonne Universit\'e, 98 bis boulevard Arago, 75014 Paris, France
\and
Laboratoire de Physique Subatomique et de Cosmologie, Universit\'e Grenoble Alpes, CNRS/IN2P3, 53 Avenue des Martyrs, Grenoble, France
\and
Consejo Superior de Investigaciones Cient\'{\i}ficas, E-28006 Madrid, Spain
\and
Universidad de Cantabria, Departamento de Ingeniería de Comunicaciones, Edificio Ingenieria de Telecomunicación, Plaza de la
Ciencia 1, 39005 Santander, Spain
\and
Astrophysics Group, Cavendish Laboratory, University of Cambridge, J J Thomson Avenue, Cambridge CB3 0HE, UK
\and
Kavli Institute for Cosmology, University of Cambridge, Madingley Road, Cambridge CB3 0HA, UK
\and
Instituto de F\'{\i}sica de Cantabria (IFCA), CSIC-Univ. de Cantabria, Avda. los
Castros, s/n, E-39005 Santander, Spain
\and
Departamento de F\'{\i}sica Moderna, Universidad de Cantabria,
Avda. de los Castros s/n, 39005 Santander, Spain
\and
CNRS-UCB International Research Laboratory, Centre Pierre Binétruy,
IRL2007, CPB-IN2P3, Berkeley, CA 94720, USA
\and
Jodrell Bank Centre for Astrophysics, Alan Turing Building, Department of Physics \& Astronomy, School of Natural Sciences, The University of Manchester, Oxford Road, Manchester, M13 9PL, U.K. 
\and
Departamento de F\'{\i}sica. Facultad de Ciencias. Universidad de
C\'ordoba. Campus de Rabanales, Edif. C2. Planta Baja.  E-14071
C\'ordoba, Spain.
\and
Department of Physics, Xi'an Jiaotong-Liverpool University, 111 Ren'ai
Road, Suzhou Dushu Lake Science and Education Innovation District,
Suzhou Industrial Park, Suzhou 215123, P.R. China. }

\date{Received 1 July 2024; accepted xx xx 2024}

  \abstract{This work focuses on the study of the Anomalous Microwave Emission (AME), an important emission mechanism between 10 and 60\,GHz whose polarization properties are not yet fully understood, and is therefore a potential contaminant for future CMB polarization observations. We use new QUIJOTE-MFI maps at 11, 13, 17 and 19\,GHz obtained from the combination of the public wide survey data and additional 1800\,h of dedicated raster scan observations, together with other public ancillary data including WMAP and \textit{Planck}, to study the polarization properties of the AME in three Galactic regions: $\rho$\,Ophiuchi, Perseus and W43. 
  
 We have obtained the spectral energy distribution (SED) for those three regions over the frequency range $0.4$--$3000$\,GHz, both in intensity and polarization. The intensity SEDs are well described by a combination of free-free emission, thermal dust, AME and CMB anisotropies. In polarization, we extracted the flux densities using all available data between 11 and 353\,GHz. We implemented an improved intensity-to-polarization leakage correction that has allowed for the first time to derive reliable polarization constraints well below the 1\% level from \textit{Planck}-LFI data. A frequency stacking of maps in the range 10--60\,GHz has allowed us to reduce the statistical noise and to push the upper limits on the AME polarization level.
  
We have obtained upper limits on the AME polarization fraction of order $\lesssim 1\%$ (95\% confidence level) for the three regions.
In particular we get $\Pi_{\rm AME} < 1.1\%$  (at 28.4\,GHz), $\Pi_{\rm AME} < 1.1\%$ (at 22.8\,GHz) and $\Pi_{\rm AME} < 0.28\%$ (at 33\,GHz) in $\rho$\,Ophiuchi, Perseus and W43 respectively. At the QUIJOTE 17\,GHz frequency band, we get $\Pi_{\rm AME}< 5.1\%$ for $\rho$\,Ophiuchi, $\Pi_{\rm AME}< 3.5\%$ for Perseus and $\Pi_{\rm AME}< 0.85\%$ for W43.  We note that for the $\rho$\,Ophiuchi molecular cloud, the new QUIJOTE-MFI data have allowed to set the first constraints on the AME polarization in the range 10--20\,GHz.
Our final upper limits derived using the stacking procedure are $\Pi_{\rm AME} < 0.58\%$ for $\rho$\,Ophiuchi, $\Pi_{\rm AME} < 1.64\%$ for Perseus and $\Pi_{\rm AME} < 0.31\%$ for W43. Altogether, these are the most stringent constraints to date on the AME polarization fraction of these three star-forming regions.}


\maketitle
\nolinenumbers 

\section{Introduction} 
\label{sec: Introduction}
The characterization of polarized Galactic foregrounds \citep{Ichiki_14} in the microwave and sub-millimetre ranges is fundamental to search for the inflationary B-mode anisotropy in the Cosmic Microwave Background (CMB) polarization \citep{kamionkowski1997, Zaldarriaga1997}. This B-mode signal, generated by inflationary gravitational waves, is contaminated by Galactic foregrounds. An accurate modelling of these foregrounds becomes very important to produce clean CMB maps suitable for their cosmological exploitation, both in intensity and in polarization. Synchrotron and thermal dust emissions are known to be strongly polarized. The former is generated by cosmic rays spiralling in the Galactic magnetic field and is known to have polarization fractions of up to $\sim 40\%$ \citep{kogut2007}, while the latter is originated in the Galactic interstellar dust and has polarization fractions of up to $\sim 20\%$ in some regions of the sky \citep{Planck2015XIX,planckX,Planck2015XXV}.
The free-free emission from thermal bremsstrahlung is known to have practically zero polarization. While the mechanisms responsible for synchrotron, thermal dust and free-free emissions are physically well understood, there is a fourth important Galactic foreground, coined as ``Anomalous Microwave Emission'' (AME) whose nature and polarization properties are still under debate. The first evidence of Galactic AME was achieved about 25 years ago as a dust-correlated signal at frequencies 10-60\,GHz that could not be explained in terms of other physical mechanisms \citep{kogut1996, leitch1997}. Neither free-free nor synchrotron were able to explain the AME observed properties. Its spectrum, characterized by a bump peaking at $\sim 20-30$\,GHz and being notably different from those of free-free and synchrotron emissions, suggested a scenario with a fresh new component emission, important through the 10-60\,GHz frequency range \citep{deOliveira1999, Watson2005, Hildebrandt2007}. 

Significant efforts have been dedicated over the last years to improve the observational characterization of AME in intensity and in polarization, with the goal to shed light on theoretical models. Observations of large sky areas \citep{deOliveira1998, deOliveira1999, davies2006, kogut2007, Todorovic2010, Macellari2011, Planck2015XXV, Rennie2022, ameplanewidesurvey}, of individual Galactic clouds \citep{Watson2005, Casassus2006, Dickinson2009, AMI2009, Tibbs2010, Vidal2011, planck2011XXnewlight, planck2015galacticcloudsAME,AMEwidesurvey}, deriving constraints in some cases on the AME polarization degree \citep{battistelli2006, Dickinson2006, Casassus2007, casassus2008, mason2009, rgspleiades, Perseus, W44, battistellispvariations,Taurus}, and of extra-galactic objects \citep{Murphy2010, NGC6946AME2, peelAMEgalaxies, planck2015galacticcloudsAME, Hensley2015, NGC4125AME, planckM33, m31AME, Linden2020,Bianchi2022, M31quijotemfi} have contributed to the understanding of the physical properties of this emission. Determining if the AME presents any polarization level is of vital importance for missions searching for the faint B-mode signal \citep{SO_19,S4_22,PTEP_LiteBIRD}. As demonstrated by \cite{remazeilles2016}, neglecting an AME component with a polarization fraction as low as $\sim 1\%$ could potentially lead to a non-negligible bias on the measured tensor-to-scalar ratio. 

Different models and theories have been proposed to explain the origin of AME. Probably the most accredited model is the electric dipole emission from small fast-spinning dust grains in the interstellar medium (ISM) \citep{draine1998a, draine1998b,spdust1, Hoangh2010, ysard2011, spdust2, ali-hamoud_review, Hoang2013, ysard2022}. There are two main hypothesis regarding the exact composition of these dust grains: the first one suggests that polycyclic aromatic hydrocarbons (PAHs) could be  responsible for this signal excess \citep{Erickson1957, draine1998a, draine1998b}, on the basis of the correlation between AME and mid-infrared dust emission in PAH-dominated bands at 8-12\,$\mu$m \citep{Ysard2010}; the second theory suggests that generic very small grains (VSG) could generate this emission \citep{hensley2016, hensley2017}. Unfortunately the exact shape of the spinning dust spectra depends on a large number of parameters that are not sufficiently well constrained observationally, thus complicating the confirmation of any of the models by observations \citep{spdust1, ysard2011, ali-hamoud_review}. 

A different model known as magnetic dipole emission (MDE) has also been proposed. In this case a magnetic field produces the alignment of the grains so they emit radiation when their minimum energy state is reached. Differently from spinning dust, MDE is a mechanism of thermal emission \citep{draine1999magnetic, drainehensley2013magneticpolarization}. This scenario seems to be disfavoured by the current upper limits on the AME polarization fraction, which are at a level  of $\lesssim 1 \%$  \citep{caraballo2011, Dickinson2011, JARM2012, W44}, while most models of MDE predict higher values \citep{draine1999magnetic, drainehensley2013magneticpolarization, hoanglazarian2016magneticpolarization}. However, \cite{draine1999magnetic} also proposed a model with random inclusions of metallic Fe that produces very low polarization (< 1\%). An alternative model based on thermal emission from amorphous dust grains is also able to reproduce the AME microwave bump in total intensity \citep{Jones2009, nashimoto2020}. For a more detailed and complete review on models and observational status of AME, see \cite{Dickinson2018}. 

In this paper we present a detailed analysis, in intensity and in polarization, of the AME in three of the brightest and best-studied Galactic regions: the $\rho$\,Ophiuchi and Perseus molecular clouds and the W43 molecular complex. $\rho$\,Ophiuchi and Perseus are ideal sources for the study of AME because they are located in regions with relatively low Galactic emission and also because they have a very low level of free-free emission, therefore enabling a clean separation of the AME component. On the other hand, W43 has significant free-free emission, but is amongst the Galactic regions harbouring more AME.
The main novelty of this work is the study of these three regions with a new and more sensitive dataset at frequencies sensitive to AME. We used new maps of QUIJOTE-MFI at 10-20\,GHz obtained through a combination of wide-survey data covering the full northern sky \citep{mfiwidesurvey} and deeper and more sensitive observations of these sources. 
The paper is organized as follows. Section~\ref{sec:regions} presents a brief description of the physical properties of the three studied regions. Section \ref{sec: DATA} describes the data set used to build the intensity spectral energy distributions (SEDs) and to derive the polarization constraints. Section \ref{sec: METHODOLOGY} describes the methodology used, including the aperture photometry technique to extract flux densities and the component-separation via modelling of the derived spectral energy distributions (SEDs) using a Markov Chain Monte-Carlo (MCMC) technique. We also describe in this section the colour-correction methodology, the correction of the intensity-to-polarization leakage in \textit{Planck}-LFI and a frequency-stacking technique aimed at improving the AME polarization constraints. Section~\ref{sec:results} presents our main results obtained on $\rho$\,Ophiuchi, Perseus and W43. The main conclusions of this work are presented in Section~ \ref{sec:conclusions}. 

\section{The Galactic regions $\rho$\,Ophiuchi, Perseus and W43} \label{sec:regions}
\begin{figure*}[!h]
\begin{center}
\includegraphics[trim= 0mm 18mm 0mm 8mm,width=8.8cm]{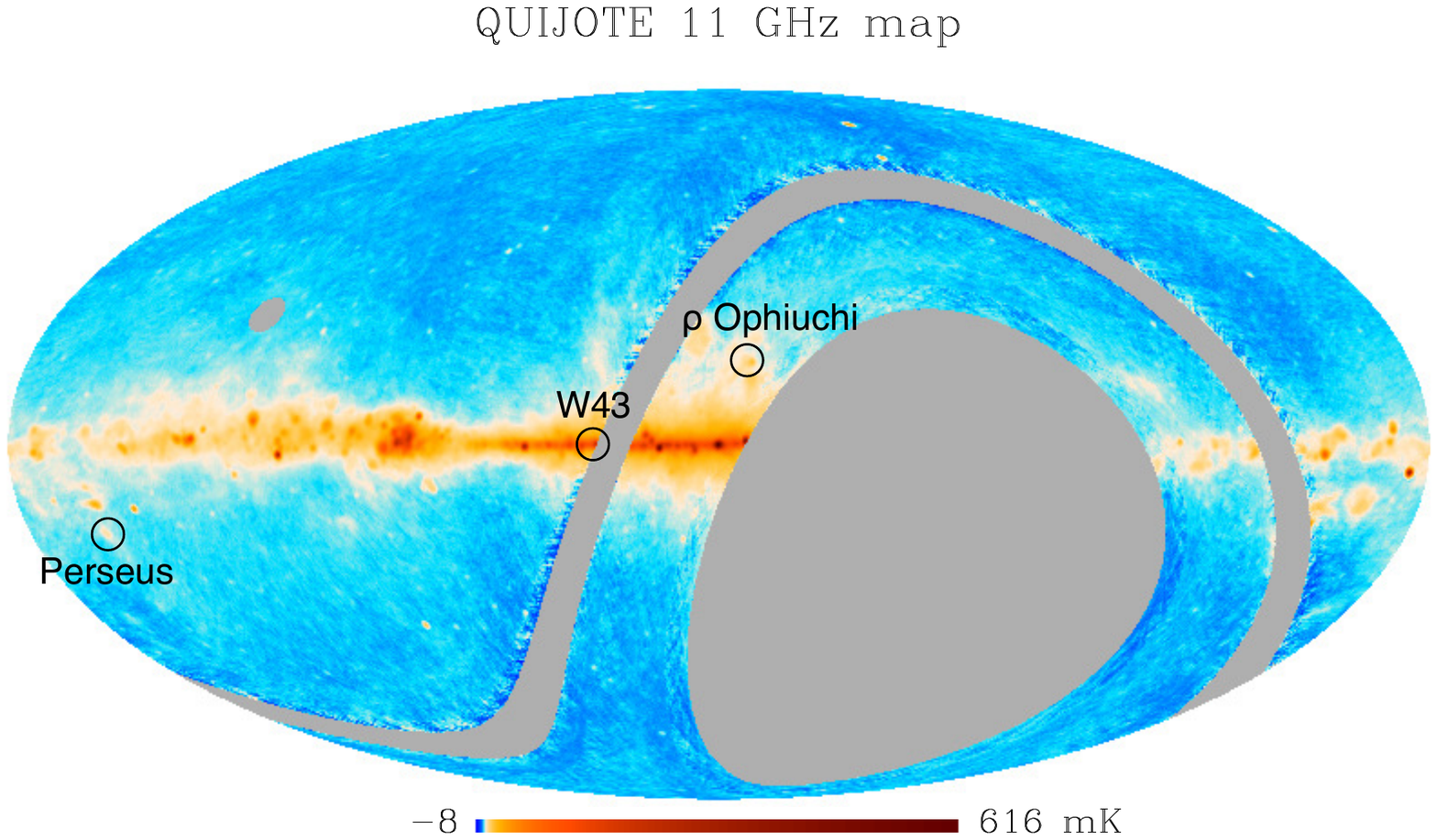}\hspace{0.2cm}
\includegraphics[trim= 0mm 18mm 0mm 8mm,width=8.8cm]{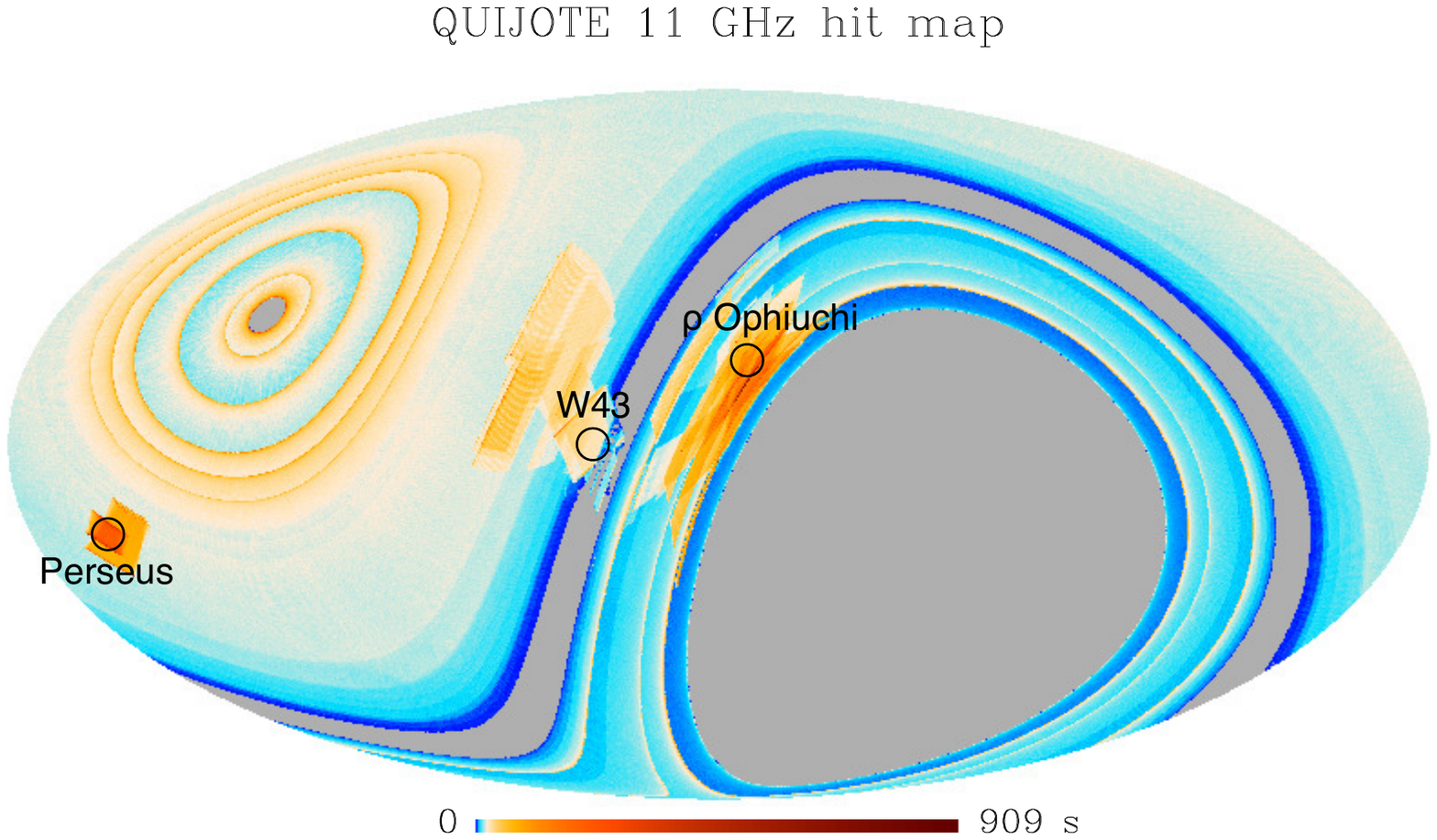}
\caption{Left: QUIJOTE-MFI wide survey intensity map at 11 GHz \citep{mfiwidesurvey}, with the locations of the three regions studied in this paper overlaid. Right: map of number of hits (per pixel of {\tt HEALPix} $N_{\rm side}=512$ and in units of seconds), for horn 3 11\,GHz, after combination of the wide survey data in nominal mode with the raster-scan data listed in Table~\ref{tab:obs}.}
\label{fig: generalmap}
\end{center}
\end{figure*}

In this section we present a brief description of the physical properties of the three Galactic regions that are the focus of this work: the $\rho$\,Ophiuchi and Perseus molecular clouds and the W43 molecular complex. Left panel of Figure~\ref{fig: generalmap} shows the location on the sky of these three sources, superimposed on the QUIJOTE 11\,GHz wide survey map. Their central coordinates, which have been taken from the SIMBAD database\footnote{\tt http://simbad.u-strasbg.fr/simbad/}, are listed in  Table~\ref{tab: AP_configurations}. Figure~\ref{fig:regions_planck857} displays high angular-resolution maps of \textit{Planck}-HFI 857\,GHz showing the different substructures of these regions.

\subsection{$\rho$\,Ophiuchi molecular cloud}
\label{sec: rho Ophiuchi}
\begin{figure*}[!h]
\begin{center}
\includegraphics[trim= 0mm 10mm 0mm 0mm, width=5.5cm]{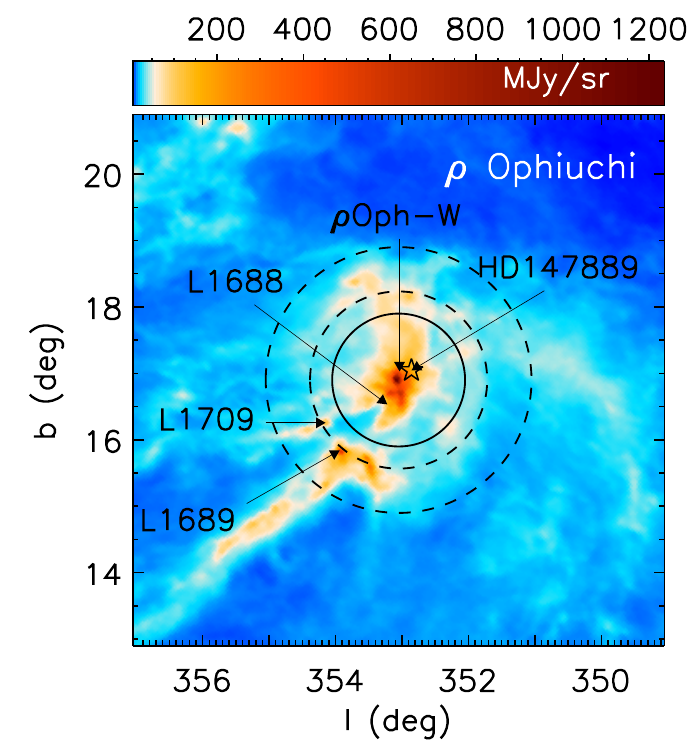}
\includegraphics[trim= 0mm 10mm 0mm 0mm, width=5.5cm]{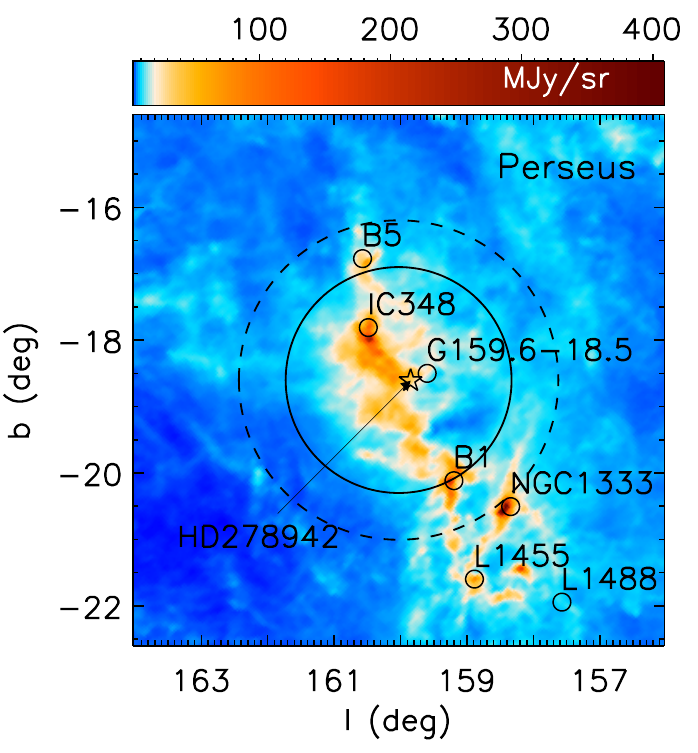}
\includegraphics[trim= 0mm 10mm 0mm 0mm, width=5.5cm]{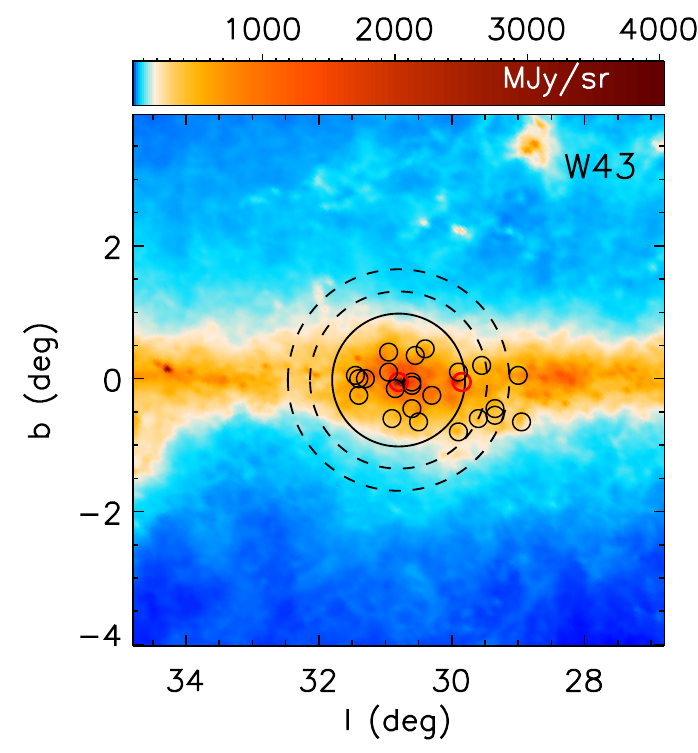}
\end{center}
\caption{High-angular resolution maps from Planck-HFI 857\,GHz around the positions of the three studied regions. In the case of the $\rho$\,Ophiuchi and Perseus molecular clouds we indicate the positions of different compact clouds extracted from different catalogues and the location of the main ionizing star. In W43 we indicate the positions of the molecular clouds identified in the CO survey of \cite{Solomon1987}, highlighting in red the two most massive ones. The solid circle delineates the aperture used for flux density integration and the dashed circles enclose the ring used for background subtraction (see Sect.~\ref{subsec: photometry}).}
\label{fig:regions_planck857}
\end{figure*}

$\rho$\,Ophiuchi is a molecular cloud in the Gould Belt located around $\sim 1^\circ$ south of the $\rho$\,Ophiuchi star, with an angular size $\approx 5^\circ$. At distance of \mbox{$D= 144\pm 7$\,pc} \citep{Zucker_19} it is the closest star-forming region to Earth. It is undergoing intermediate star formation, concentrated in three clouds of dense gas and dust: the Lynds dark clouds L\,1688, which contains the Ophiuchus star cluster and is considered the main cloud of this complex \citep{Abergel1996}, L\,1689 and L\,1709 (see Figure~\ref{fig:regions_planck857}). Ultra-violet radiation from the hottest young stars in this cluster dissociates the surrounding gas. The best example is the prominent photodissociation region (PDR) $\rho$\,Oph-W that is excited by the star B2V HD147889 and constitutes the western edge of L1688 \citep{liseau1999,Habart2003}. This is the region where the bulk of the AME is produced. This was first identified by \cite{casassus2008} as an excess of emission at 31\,GHz using data from the CBI interferometer. AME in this region was subsequently studied by \cite{Dickinson2011}, who derived upper limits on its polarization fraction of the order of $\lesssim 1\%$, and by \cite{planck2011XXnewlight}. More recently, \cite{Arce-Tord_20} discovered spatial variations on the spinning dust emissivity using observations of the CBI2 interferometer, while \cite{Casassus_21} used observations with ATCA, at a finer angular resolution, to study the AME in this region at smaller scales.

\subsection{Perseus molecular cloud}
\label{sec: perseus}
The Perseus molecular cloud complex is a relatively nearby giant molecular cloud at a distance of $294\pm 15$\,pc \citep{Zucker_19}. The full cloud is around 30\,pc across ($\sim 6^\circ\times 3^\circ$ on the sky) and encompasses six dense cores: B\,5, IC\,348, B\,1, NGC\,1333, L\,1455 and L\,1448 (see Figure~\ref{fig:regions_planck857}). AME originates mainly around the dust shell G159.6-18.5 located southwest of IC348, that is illuminated by the O9.5-B0V star HD278942, and filled by an HII region \citep{Andersson_00}. AME from G159.6-18.5 was first detected by \cite{Watson2005} using data from the COSMOSOMAS experiment, a result that is widely recognised as the first unambiguous detection of AME in a compact region. This region dominated most of the dust-correlated signal first identified by \cite{deOliveira1999} via correlations between data at 10\,GHz and 15\,GHz from the Tenerife experiment and dust maps. Using high-angular resolution data at 33\,GHz with the VSA interferometer, \cite{Tibbs2010} concluded that the bulk of the AME is diffuse (originated in scales larger than 10\,arcmin, that is the angular resolution of the VSA). \cite{battistelli2006} analyzed 11\,GHz data in polarization from the COSMOSOMAS experiment and found a tentative signal with a polarization fraction of $3.4^{+1.5}_{-1.9}\%$, whereas \cite{caraballo2011} and \cite{Dickinson2011} determined upper limits of $\lesssim 1\%$ (95\% C.L.) on the AME polarization fraction using WMAP 23\,GHz data\footnote{Note that \cite{caraballo2011} quote polarization upper limits with respect to the total measured intensity emission, while \cite{Dickinson2011} use the residual AME intensity emission, which is the same we do in this work}. More recently \cite{Perseus} presented new flux densities and polarization upper limits using QUIJOTE MFI commissioning data with a shallower sensitivity than those used in this paper. \cite{planckemissivityestimates} applied a different analysis consisting of looking for correlations between a weighted polarized intensity map constructed from the combination of WMAP and \textit{Planck} data and the AME intensity map from {\tt Commander}, on a larger region around the Perseus molecular cloud, to derive an upper limit of $<1.6\%$.

\subsection{W43 molecular complex}
\label{sec: Ws}
W43 (source number 43 of the catalogue of \citealt{Westerhout_58}) is one of the richest molecular complexes and with one of the highest star formation rates in our Galaxy \citep{NguyenLuong2011}. It is located at a distance of $\approx 5.5$\,kpc and has a physical size of $\sim 140$\,pc, extending almost $2^\circ$ along the direction of Galactic longitude.  According to \cite{NguyenLuong2011} this complex includes more than 20 molecular clouds with high velocity dispersion \citep{Solomon1987} and is surrounded by atomic gas that extends up to $\sim 290$\,pc. In Figure~\ref{fig:regions_planck857} we show the locations of these compact molecular clouds, highlighting (red circles) the positions of W43-main and W43-south that are the most massive ones \citep{NguyenLuong2011}. The core of W43-main harbours a well-known giant HII region powered by a particularly luminous cluster of Wolf-Rayet and OB stars \citep{Blum1999}. AME in W43 was first identified by \cite{Irfan2015}. Using new data from QUIJOTE MFI, \cite{W44} determined an upper limit on the AME polarization fraction of $<0.22\%$ that, as of today, is the most stringent constraint on the polarization of the AME. These results are revisited in this paper.

\begin{table}
\caption{Basic characteristics of the sources studied in this paper. The name and the physical type (photodissociation region or molecular cloud) are indicated in the first two columns. Central coordinates are shown in the next two columns. The last three columns show the radii of the aperture and of the background ring that are used in section~\ref{sec: METHODOLOGY} to extract flux densities.}
\begin{center}
\begin{tabular}{cccccccc}
        \hline
        & & & & \multicolumn{3}{c}{Aperture parameters}  \\
	
	Source             & Type   & l            & b            & $\theta_{\rm ap}$   & $\theta_{\rm int}$    & $\theta_{\rm ext}$     \\
                           &        & ($^{\circ}$)	& ($^{\circ}$)  & (')   & (')     & (')      \\
	\hline
        $\rho$\,Ophiuchi& PDR    & 353.05       & 16.90        & 60         & 80           & 120           \\
        Perseus            & MC       & 160.03       & $-18.6$         & 102        & 102          & 144           \\
	W43                & MC    & 30.8         &  $-0.02$ 	  & 60         & 80           & 100 	       \\
	\hline
	
\end{tabular}
\end{center}
\label{tab: AP_configurations}
\end{table}

\section{Data} \label{sec: DATA}
We used twenty five total-intensity maps between 0.408 GHz and 3000 GHz to build the SEDs of the three regions, and sixteen maps in polarization. In Table~\ref{tab: surveys summary} we list the main properties of these maps. Although we indicate the parent angular resolution of these maps, all of them have been smoothed to an effective angular resolution of $1^\circ$. They all use a {\tt HEALPix} \citep{Healpix} pixelization with resolution $N_{\rm side}=512$. Details of each of these surveys are given in the following subsections.
\begin{table*}
\caption{The maps used in this paper, including central frequency, calibration error, angular resolution (beam full-width half maximum), covered sky fraction, an indication of whether or not there is polarization information and reference.}
\begin{center}
\begin{tabular}{ccccccc}
	\hline
	Name 	        		& Freq.    & Calibration error       & FWHM  	    & Sky Coverage 	    & Polarization		        & References                                \\
	 	        		& (GHz)    &  (\%)                   & (arcmin) 	&  	                & 		                    &                                           \\
	\hline	
	Haslam		 	        & 0.408 		& 10                    & 52   			    & All-sky		                        & No	   			& \cite{haslam1982}	    \\
	Dwingeloo			    & 0.82 			& 10                    & 72	     		& $\delta$ > -7$^{\circ}$			    & No   			    & \cite{dwingeloo}      \\
	Reich			        & 1.42 		& 10               	    & 36     			& All-sky 	                            & No                &\cite{effelsberg1}, \cite{effelsberg2}\\
        S-PASS			        & 2.3			& 10                    & 8.9			    & $\delta$ < 1$^{\circ}$			    & Yes				& \cite{carretti2019}	\\
	HartRAO 			    & 2.3 		& 10                    & 20    			& $\delta$ < 13$^{\circ}$			    & No  			    & \cite{hartrao}		\\
	QUIJOTE-MFI		        & 11.1			& 5                     & 53.2				& $\delta$ > -32$^{\circ}$			    & Yes			    & \cite{mfiwidesurvey}  \\
	QUIJOTE-MFI		        & 12.9			& 5                     & 53.5				& $\delta$ > -32$^{\circ}$			    & Yes			    & \cite{mfiwidesurvey}	\\
	QUIJOTE-MFI		        & 16.8			& 5                     & 39.1			    & $\delta$ > -32$^{\circ}$			    & Yes			    & \cite{mfiwidesurvey}  \\
	QUIJOTE-MFI		        & 18.7			& 5                     & 39.1			    & $\delta$ > -32$^{\circ}$			    & Yes			    & \cite{mfiwidesurvey}  \\
	WMAP K-band 		    & 22.8 			& 3                     & 51.3  			& All-sky			                    & Yes		    	& \cite{wmap}			\\
	\textit{Planck} LFI     & 28.4 			& 3                     & 33.1    			& All-sky	    	                    & Yes  			    & \cite{planck2018cmb}	\\
	WMAP Ka-band 	        & 33.0 			& 3                     & 39.1    			& All-sky		                        & Yes  			    & \cite{wmap}		    \\
	WMAP Q-band		        & 40.6			& 3                     & 30.8   			& All-sky		                        & Yes			    & \cite{wmap}			\\
	\textit{Planck} LFI     & 44.1 			& 3                     & 27.9  			& All-sky		                        & Yes		    	& \cite{planck2018cmb}	\\
	WMAP V-band		        & 60.4			& 3                     & 21.0	     		& All-sky		                        & Yes  			    & \cite{wmap}	        \\
	\textit{Planck} LFI     & 70.5 			& 3                     & 13.1      		& All-sky		                        & Yes  			    & \cite{planck2018cmb}	\\
	WMAP W-band         	& 93.5 			& 3                     & 14.8     			& All-sky		                        & Yes  		    	& \cite{wmap}			\\
	\textit{Planck} HFI     & 100 			& 3                     & 9.7  			    & All-sky		                        & Yes 			    & \cite{planck2018cmb} 	\\
	\textit{Planck} HFI 	& 143 			& 3                     & 7.3 	     		& All-sky		                        & Yes   			& \cite{planck2018cmb}	\\
	\textit{Planck} HFI		& 217		    & 3                     & 5.0     			& All-sky		                        & Yes   			& \cite{planck2018cmb}	\\
	\textit{Planck} HFI     & 353 			& 3                     & 4.9   			& All-sky		                        & Yes  			    & \cite{planck2018cmb}	\\
	\textit{Planck} HFI		& 545 			& 6.1                   & 4.8   			& All-sky		                        & No			    & \cite{planck2018cmb}	\\
	\textit{Planck} HFI     & 857 			& 6.4                   & 4.6     			& All-sky		                        & No  			    & \cite{planck2018cmb}	\\
	COBE-DIRBE  	        & 1249 			& 11.6                  & 37.1   			& All-sky		                        & No 			    & \cite{cobe-dirbe}	    \\
        COBE-DIRBE   	        & 2141			& 10.6                  & 38.0   			& All-sky		                        & No 			    & \cite{cobe-dirbe}	    \\
	COBE-DIRBE  	        & 2997 		    & 13.5                  & 38.6   			& All-sky		                        & No 			    & \cite{cobe-dirbe}	    \\
	\hline	
\end{tabular}
\end{center}
\label{tab: surveys summary}
\end{table*}
\subsection{QUIJOTE data} 
\label{subsec: QUIJOTE data}
The new data presented in this paper were acquired with the QUIJOTE experiment, \citep{RubinoSPIE12}. One of the science drivers of this experiment is to characterize the polarization of the low-frequency foregrounds, mainly the synchrotron and the AME. QUIJOTE is located at the Teide Observatory (Tenerife, Spain) at 2400\,m above the sea level and at geographical longitude $16^\circ 30' 38''$ West and latitude $28^\circ 18' 04''$ North. Observing at the minimum elevation attainable by QUIJOTE of $30^\circ$ at this latitude allows reaching declinations as low as $-32^{\circ}$. QUIJOTE consists of two telescopes with an offset crossed-Dragone optics design, with projected apertures of $2.25$\,m for the primary and $1.89$\,m for secondary mirror, providing highly symmetric beams (ellipticity $< 0.02$) with very low sidelobes ( $\leq$ 40\,dB) and polarization leakage ($\leq$ 25\,dB). This optics and mount were chosen to allow the telescope to spin fast at a constant elevation while observing. The two telescopes are equipped with three instruments covering the frequency range $10-40$\,GHz. The first instrument on the first QUIJOTE telescope, the so-called Multi-Frequency Instrument (MFI), consisted of four horns, each of which observes in 2 frequencies bands: horns 1 and 3 observe at 11 and 13\,GHz, while horns 2 and 4 at 17 and 19 GHz, each with a 2\,GHz bandwidth. The full width at half-maximum (FWHM) is $\approx 55$\,arcmin at 11 and 13\,GHz, and $\approx 39$\,arcmin at 17 and 19 GHz (G\'enova-Santos et al. in prep.). The data used in this paper were taken with this instrument.
\subsubsection{New raster-scan observations}
\label{subsubsec:observation}
The QUIJOTE-MFI instrument observed between 2012 and 2018. Most of the time during this period (more than 9000 hours) was dedicated to observations in the so-called ``nominal mode'' (continuous rotation of the telescope at constant elevation), leading to maps covering the full northern sky (total sky fraction of $\approx 73$\%) and with sensitivities of $\sim 60-200$ $\mu$K deg$^{-1}$ in intensity and $\sim 35-40$ $\mu$K deg$^{-1}$ in polarization . These ``wide survey'' maps were publicly released in January 2023 and their properties are described in detail in \cite{mfiwidesurvey}. This paper uses a combination of these data in the nominal mode with deeper observations in raster-scan mode, leading to higher sensitivities at the positions of these regions.

The QUIJOTE-MFI raster-scan observations consisted of back-and-forth constant-elevation scans of the telescope performed with an effective scanning speed on the sky of 1\,deg/s (the telescope is moved with angular velocity around the azimuth axis $\omega_{\rm AZ}=$1/cos(EL)\,deg/s). Each observation was typically comprised of a few hundred scans\footnote{We define a scan as the movement of the telescope at a fixed elevation between two fixed azimuths, either westwards or eastwards.} (total duration per observation of $\sim 1$\,hour), in such a way that rotation of the sky leads to a map size along the elevation direction similar to the scan length along the azimuth direction. Typically between one and five observations were performed every day, and were repeated in consecutive days with a civil time offset of 4~minutes (same sidereal time). Table \ref{tab:obs} presents a summary of the observations in raster-scan mode that are used in this paper, including total integration times. Leaving aside the observations in the nominal mode leading to the wide survey maps, these fields, and particularly HAZE and PERSEUS, are amongst the fields with the highest total observing time of QUIJOTE-MFI. The final maps of $\rho$\,Ophiuchi combine observations in this field with wider observations in the fields HAZE and HAZE2 intended to investigate the excess of microwave emission around the Galactic Centre that has been addressed in \cite{hazewidesurvey}. The HAZE and HAZE2 observations are clearly reflected in the map of number of hits of Figure~\ref{fig: generalmap} as a redder wide region south of the $\rho$\,Ophiuchi field. The redder region to the northeast of W43 corresponds to the HAZE3 field, which has not been included in Table~\ref{tab:obs} because it does not overlap with either of the three regions that we study in this paper. The $\rho$\,Ophiuchi maps used in this paper are the same as in \cite{hazewidesurvey}. 

We have performed three different types of observations around the Perseus molecular cloud, as indicated in Table~\ref{tab:obs}. The so-called PERSEUS field consists of azimuth scans of size $15^\circ$. This value is close to the minimum scan size in QUIJOTE-MFI observations so that the source is observed by the four horns in a single observation. In order to maximize the integration time per unit solid angle, and therefore to improve the map sensitivity, in this case we also performed the observations called PERSEUS-H2 and PERSEUS-H3 that are respectively centred in horns 2 and 3 and use a smaller scan length of $5^\circ$ and $6^\circ$ respectively. Given the smaller map size, in these cases the source is only seen by horn 2 in PERSEUS-H2 and by horn 3 in PERSEUS-H3. This observing strategy leads to much higher integration time per unit solid angle (see values in Table~\ref{tab:obs}). Note that these are a different set of observations from those used in \cite{Perseus} that were performed between December 2012 and April 2013 during the commissioning of QUIJOTE-MFI. In the final Perseus maps presented here we have discarded those observations because at that time the internal calibration signal that is now used by default to monitor and correct gain variations (see section 2.2.1 of \citealt{mfiwidesurvey}) was not available. 

The observations in raster-scan mode in W43 were described in \cite{W44}. In this paper we use these same observations, but with an improved data processing (see subsection~\ref{sec:data_reduction}), in combination with the wide survey data presented in \cite{mfiwidesurvey}. These latter data have an average integration time per solid angle of 0.16\,h$\,$deg$^{-2}$ (see Table~\ref{tab: sensitivity}) and then will not have a significant impact on the final map sensitivities. However they help to reduce various systematic effects, and in particular the combination of more scanning directions contributes to a more efficient destriping procedure and to minimise the large-scale systematic effects in polarization. 
\begin{table*} 
\caption{Main parameters of the raster scan observations in each field. We list the periods during which these observations were done (see definition of period index in \citealt{mfiwidesurvey}, central coordinates, number of observations, length of the azimuth scan, elevations of the observations, total covered sky area, total integration time, and integration time per unit solid angle of one square degree calculated around the central position and using horn 3 as reference.}
\begin{center}
\begin{tabular}{lccccccccc} 
\hline
\smallskip
Field & Dates & Period & $n_{\rm obs}$ & ($l$,$b$) & $\Delta$AZ$\cdot$cos(EL) & EL & Area & \multicolumn{2}{c}{$t_{\rm int}$}\\
\cline{9-10}
& & & & (deg) & (deg) & (deg) & (deg$^2$) & (h) & (h\ deg$^{-2}$) \\
\hline
$\rho$\,Ophiuchi & Dec. 2015 - Dec. 2017  & 3-6  & 186 & ($353.0,16.9$)  & $14,15,25$ & $32,34,37$	&  414 & 246  &  0.85 \\
HAZE            & Aug. 2013 - Oct. 2016  & 1-5 & 328 & ($8.6,2.4$)     & $30,40,43$ & $33,37,39$	& 1509 & 719  &  0.31 \\
HAZE2           & Jul. 2014 - Aug. 2018  & 2-6  & 100 & ($357.1,22.6$)  & $25,28,30$ & $32,36,37$	&  530 &  96  &  0.41 \\
\hline
PERSEUS         & Jul. 2015 - Sep. 2015 & 2  & 149 &  ($160.2,-18.5$)& $15$	 & $33,42,51$	&  150 &  98  &  0.68 \\
PERSEUS-H2      & Oct. 2013 - Jan. 2015 &  1,2  & 432 & ($160.2,-18.5$) & $6$	 & $32-84$	&   54 & 243  &  3.92 \\
PERSEUS-H3      & Oct. 2013 - Sep. 2014   &1,2 & 404 & ($160.2,-18.5$) & $5$	 & $36-81$	&   57 & 213  &  4.00 \\ 
\hline
W43             & Mar. 2015 - Jun. 2015  & 2  & 305 &($34.7,-0.4$)    & $11,22,25$ & $36-63$ 	&  363 & 210  &  0.93  \\
\hline
\end{tabular}
\end{center}
\label{tab:obs}
\end{table*}

\subsubsection{Data reduction}\label{sec:data_reduction}
	
The QUIJOTE-MFI data processing pipeline is introduced in section~2.2 of \cite{mfiwidesurvey} and will be explained in depth in a dedicated paper (G\'enova-Santos et al. in prep.). The QUIJOTE-MFI maps on which the analyses presented in this paper are based were generated following the same procedure. Briefly: i) the global gain calibration is based on regular raster-scan observations of two bright radio sources, Tau\,A and Cas\,A; ii) the same observations of Tau\,A are used to calibrate the polarization direction of the detectors; iii) gain variations in long time scales are corrected using an internal calibration signal that is emitted by a thermally-stabilised diode every 30~seconds; iv) projection of the TOD data onto maps is done using a destriping algorithm called PICASSO \citep{destriper} that is an adaptation of the MADAM approach \citep{Keihanen_05} to QUIJOTE data. 

The previous study of QUIJOTE-MFI on the Perseus molecular cloud \citep{Perseus}, apart from being based on a different and less sensitive dataset, did not implement points (iii) and (iv), i.e. no gain correction was executed and the map making was based on a simpler median filter, which results in a less efficient removal of intensity $1/f$ noise and suppression of the angular scales larger than the filter size. The previous QUIJOTE-MFI study in W43 \citep{W44} used the same raster-scan data of this paper (but without the combination with the data in the nominal mode), as it was mentioned in the previous subsection. In that case the same destriping algorithm as in this paper was used. However the gain correction of point iii), which is an important improvement in the current analysis, was not applied. Another important difference with respect to those previous studies concerns the global gain calibration. In both \cite{Perseus} and \cite{W44} it was based on the Tau\,A and Cas\,A models presented in \cite{Weiland_11}. The maps used in this paper are calibrated instead using an improved model for Tau\,A that will be described in detail in G\'enova-Santos \& Rubi\~no-Mart\'{\i}n (in preparation; the model is given in equation~9 of \citealt{mfiwidesurvey}). The uncertainty of these models in the QUIJOTE-MFI frequency range is of the order of 5\,\%, which is considered to be the global calibration uncertainty of the QUIJOTE maps. In addition we have developed an improved and more-reliable method, based on a beam fitting algorithm, to extract from QUIJOTE-MFI data the reference flux density of Tau\,A that is used to calibrate the maps. These modifications lead to differences of the order of 5--10\% in the final flux densities of the sources. Given the improvements commented before on gain correction and calibration, the results presented in this paper should be deemed more reliable.

\subsubsection{Maps}\label{sec:maps}

Maps at each of the four QUIJOTE-MFI frequencies are produced from the calibrated time-ordered data using the destriping algorithm described in section~\ref{sec:data_reduction}. 
The map-making parameters (baseline length and priors on the correlated-noise parameters) are the same as those adopted for the wide-survey maps (see Table~5 in \citealt{mfiwidesurvey}).
Data affected by different systematic effects (radio interference, strong gain variations, etc) are flagged following the methodology and criteria explained in section~2.2.2 of \cite{mfiwidesurvey}. 
The post-processing of the maps (weights for the combination of channels and the filtering with the function of the declination, as described respectively in sections 2.4.1 and 2.4.2 of \citealt{mfiwidesurvey}) is also identical to the one used for the wide-survey maps. Table~\ref{tab: sensitivity} lists the effective integration times per unit solid angle used to generate the maps, calculated in a region around the central coordinates of each source indicated in Table~\ref{tab: AP_configurations}, except for W43 for which we used coordinates $l=35.8^\circ$, $b=-0.02^\circ$ to avoid the nearby masked region affected by contamination from geostationary satellites (see section~2.2.2 of \citealt{mfiwidesurvey}). Comparison of these numbers with the total observed times shown in Table~\ref{tab:obs} gives an idea of the fraction of flagged data in each case (note that the total integration times given in Table~\ref{tab:obs} are for horn 3). The region most affected by flagging is Perseus, owing to significant contamination from radio interference in many of the observations. On the other hand $\rho$\,Ophiuchi is the region least affected, and in this case we kept 64\% of the data at 11\,GHz. In all cases the amount of flagging is larger in polarization than in intensity. Table~\ref{tab: sensitivity} also shows a comparison of the integration times in the nominal mode and in the combination of nominal plus raster-scan data, highlighting the notably higher integration times achieved in the raster scans. This fact becomes also evident in the map of number of hits illustrated in the right panel of Figure~\ref{fig: generalmap}, which clearly shows a higher integration time in the regions where these three sources are located.

The maps at 11 and 13\,GHz were generated using only data from horn 3. As with other QUIJOTE-MFI papers, maps from horn 1 are not used due to having important systematic effects, in particular problems with the positioning of the polar modulator \citep{mfiwidesurvey}. At 17 and 19\,GHz the maps of $\rho$\,Ophiuchi and Perseus from horns 2 and 4 are combined through a weighted mean that uses predefined constant weights (see section~3 of \citealt{mfiwidesurvey}). In the case of the W43 field we use only maps from horn 2, as in this case the polarization maps of horn 4 seem to be affected by intensity-to-polarization leakage. Prior to that combination, intensity and polarization maps produced from the correlated and uncorrelated channels are also combined. In the case of polarization, uncorrelated channels are only used for data taken under a configuration such that the two channel outputs have correlated $1/f$ noise properties. All these details, as well as the definition of correlated and un-correlated channels, are explained in depth in \cite{mfiwidesurvey}. The noise of the lower and upper frequency bands of each horn are significantly correlated (up to 80\% in intensity) because of the use of the same low-noise amplifiers, as explained in section 4.3.3 of \cite{mfiwidesurvey}. Ideally the noise covariance between the 11 and 13\,GHz maps on the one hand, and between the 17 and 19\,GHz maps on the other, should be taken into account. However, we have verified that this has no significant impact on the results derived in this paper (differences of 3\% in the worst case on the derived model parameters), so for the sake of simplicity we have ignored this covariance term.

\begin{figure*}[!h]
\begin{center}
\includegraphics[width=18.8cm]{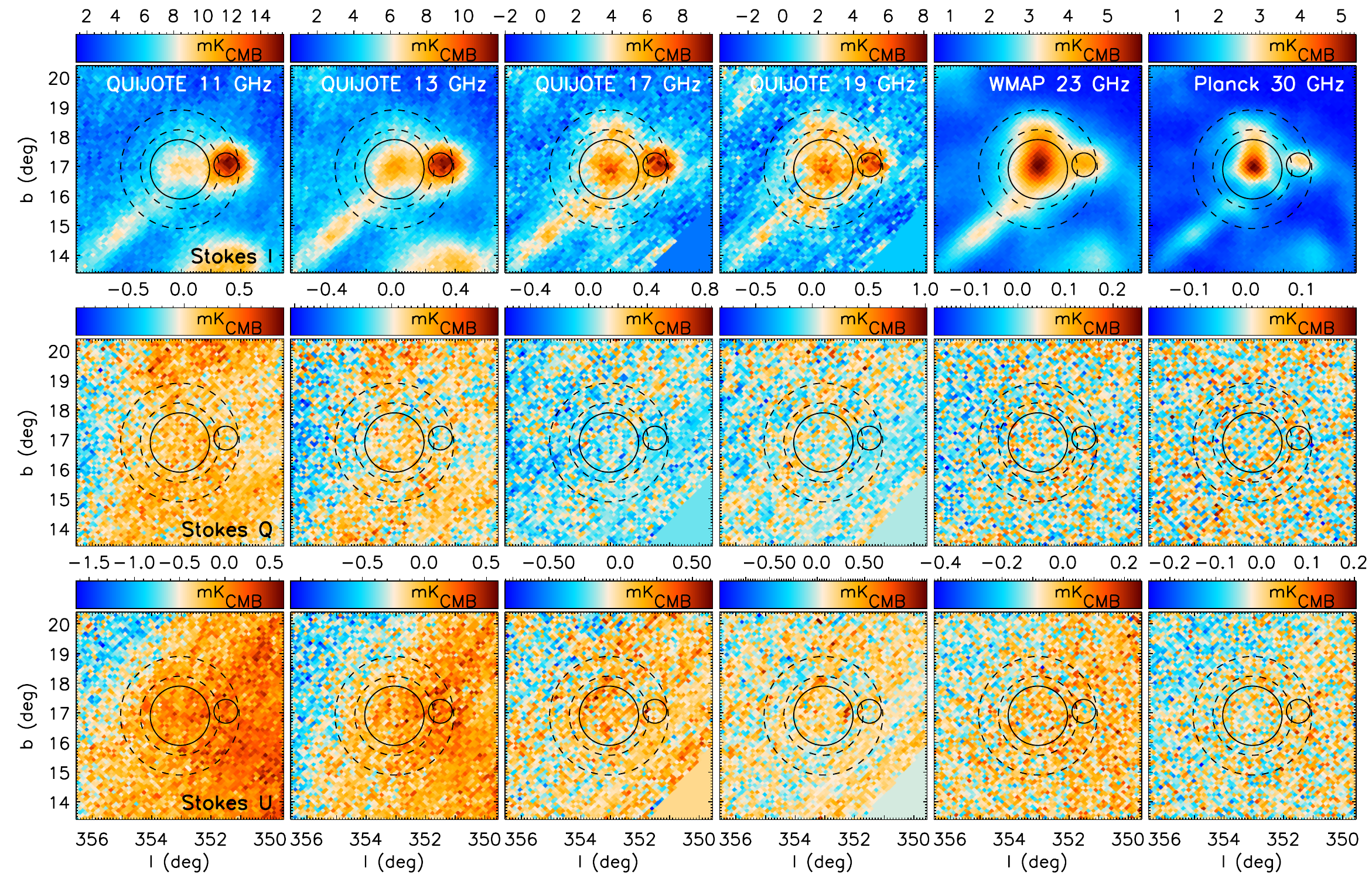}
\end{center}
\caption{Intensity and polarization maps around the $\rho$\,Ophiuchi molecular cloud from QUIJOTE-MFI and from the two lowest-frequency bands of WMAP and \textit{Planck}. The three rows show respectively $I$, $Q$ and $U$ maps while the columns correspond to 11, 13, 17 (Horn 2), 19 (Horn 2), 23 and 30\,GHz from left to right. The solid circle shows the aperture we use for flux integration whereas the two dashed circles enclose the ring we use for background subtraction. The small circle inside the background annulus toward the west indicates the mask that has been applied to avoid a strong HII source. For the sake of a better visualization these maps are shown at their raw angular resolution, although all the analyses presented in this paper have been performed on maps convolved to a common angular resolution of $1^\circ$.}
\label{fig:rhoophw_maps}
\end{figure*}

\begin{figure*}
\begin{center}
\includegraphics[width=18.8cm]{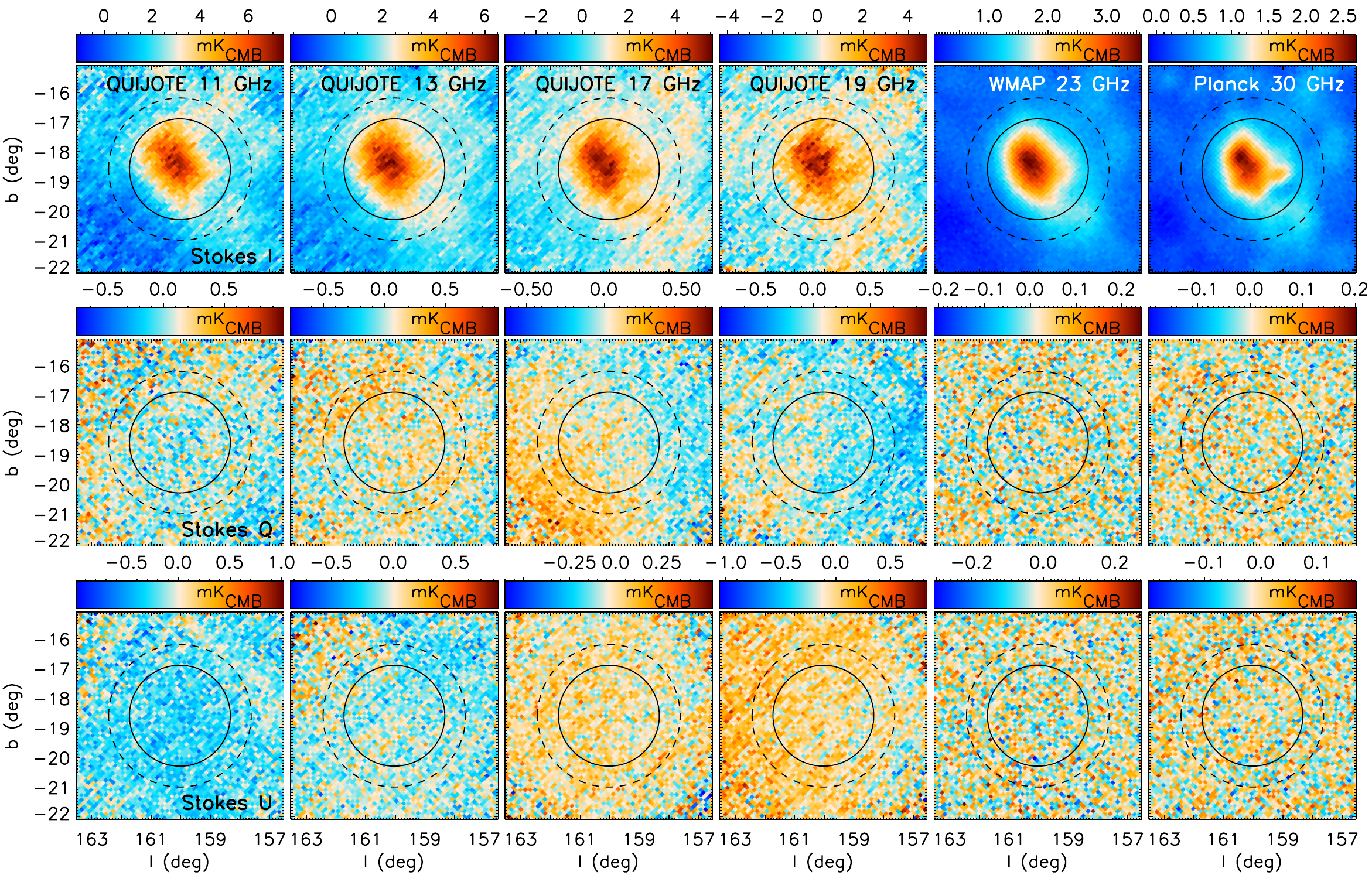}
\end{center}
\caption{Same as in Figure~\ref{fig:rhoophw_maps} but for the Perseus molecular cloud.} 
\label{fig:perseus_maps}
\end{figure*}

\begin{figure}
\begin{center}
\includegraphics[width=\columnwidth]{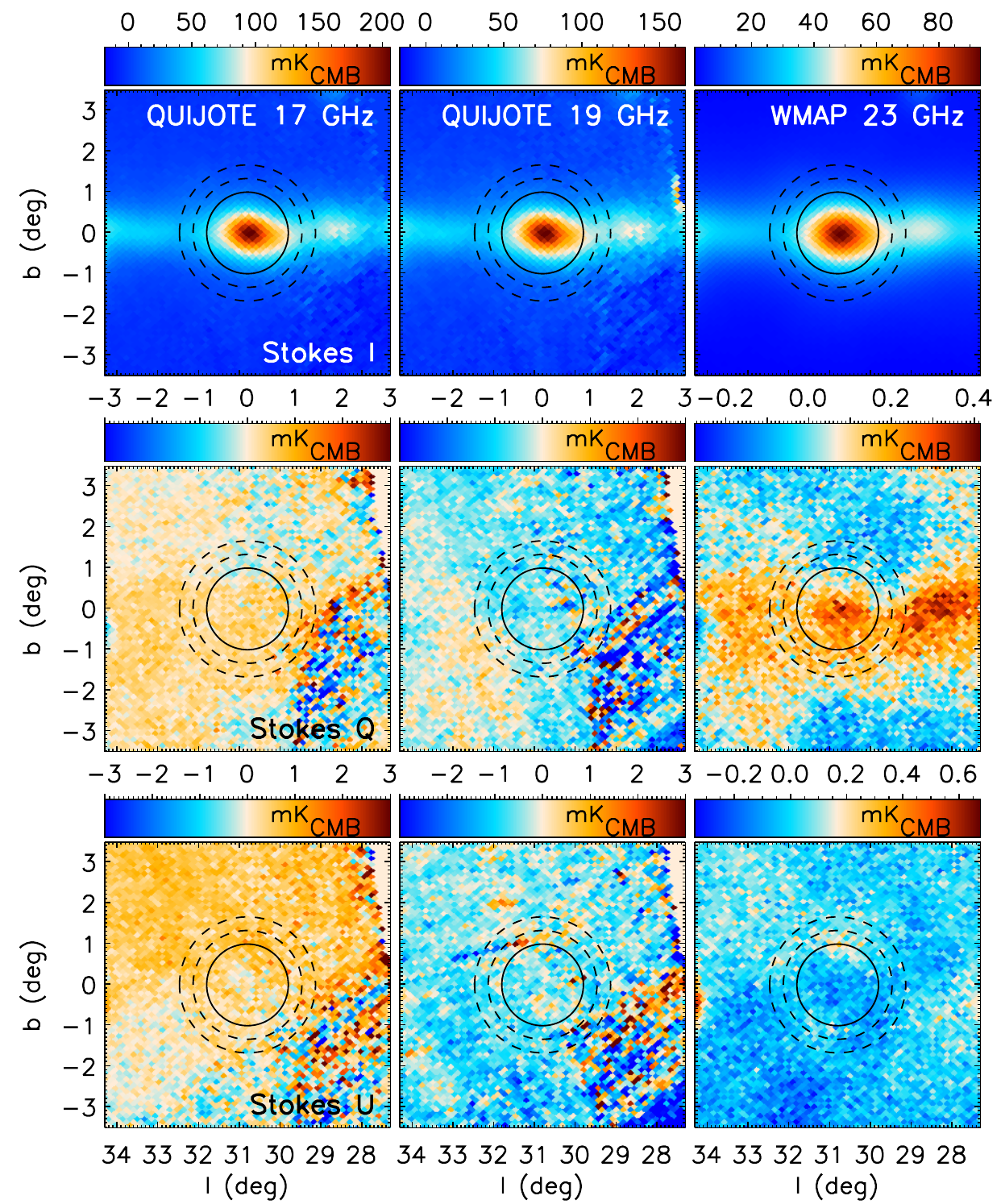}
\end{center}
\caption{Intensity and polarization maps around the W43 molecular cloud. The three rows show respectively $I$, $Q$ and $U$ maps while the columns correspond to QUIJOTE-MFI 17 (Horn 2) and 19\,GHz (Horn 2) and to WMAP 23\,GHz from left to right. The solid circle shows the aperture we use for flux integration whereas the two dashed circles enclose the ring we use for background subtraction.}
\label{fig:w43_maps}
\end{figure}

Final QUIJOTE-MFI intensity (Stokes $I$) and polarization (Stokes $Q$ and $U$) maps at their native angular resolution are shown in Figures~\ref{fig:rhoophw_maps}, \ref{fig:perseus_maps} and \ref{fig:w43_maps}, for $\rho$\,Ophiuchi, Perseus and W43, respectively. For comparison we show also the WMAP 23\,GHz and \textit{Planck} 30\,GHz maps. In total intensity these maps are clearly dominated by emission from each of these sources, and the increase of flux density from 11 to 19\,GHz associated with the AME is evident even by eye. Thanks to the presence of an adjacent HII region (its position is indicated in the figure through a solid circle), which is dominated by free-free emission, the region showing the clearest visual evidence of AME is $\rho$\,Ophiuchi. Here the photodissociation region that harbours the AME, located towards the centre of the map, becomes more and more intense relative to the free-free emission in the HII region as the frequency increases. Meanwhile, the polarization maps are mostly consistent with noise. The exceptions are: i) the diffuse signal shown at 11 and 13\,GHz in the $\rho$\,Ophiuchi maps that is due to one of the diffuse bright filaments \citep{Vidal_15} originating from the Galactic centre (see section~\ref{sec:rhoophw}), and that leaves a temperature gradient running from the northeast to the southwest, and ii) diffuse emission seen in the $Q$ map of W43 distributed along the Galactic plane that is most-likely due to diffuse Galactic synchrotron emission as already discussed in \cite{W44}. The origin of this emission is discussed in depth in section~\ref{sec:w43}, while in appendix~\ref{appendix} we present a detailed study of the possible contribution of instrumental effects to this signal.

The noise properties of these maps are evaluated from jack-knife maps resulting from the subtraction of the two half-mission maps (see section~4.1 of \citealt{mfiwidesurvey}). The noise levels in intensity and in polarization derived from these maps, in units of standard deviations in $\mu$K in a region with a solid angle of 1\,deg$^2$, are listed in Table~\ref{tab: sensitivity}. While in our analyses we use maps resulting from the combination of horns 2 and 4, as explained above, here we have quoted noise figures from these two horns independently. There is a clear improvement over the noise levels achieved in the wide survey data (nominal mode), which are of the order $30-80\,\mu$K\,deg$^{-1}$ in polarization (see Table~14 of \citealt{mfiwidesurvey}. At 11 and 13\,GHz we achieve noise levels in polarization of $\sim 7-10\,\mu$\,K\,deg$^{-1}$ in Perseus and in $\rho$\,Ophiuchi. Together with the maps obtained around the Taurus molecular cloud (see Table~1 of \citealt{Taurus}) and on M31 (see Table~2 of \citealt{M31quijotemfi}) these are amongst the deepest and most sensitive observations obtained with QUIJOTE-MFI. Instantaneous sensitivities (sensitivity in an integration of one second) per channel can be estimated from the values of Table~\ref{tab: sensitivity} as $\sigma_{\rm map}\sqrt{2t_{\rm int}}$, where $\sigma_{\rm map}$ is the map sensitivity listed in the last three columns, $t_{\rm int}$ is the integration time per unit solid angle listed under the `n+r' columns, and the factor $\sqrt{2}$ must be applied only when the maps use a combination of correlated and uncorrelated channels, so that we get the sensitivity to the measurement of $I$, $Q$ or $U$ through only one of these two combinations. This calculation gives values of the order of $0.6-1.0$\,mK\,s$^{1/2}$ in $Q$ and $U$ and of the order of $3-5$\,mK\,s$^{1/2}$ in $I$, which are consistent with the typical values derived in other regions (see e.g. Table~13 of \citealt{mfiwidesurvey}).
\begin{table}
\caption{Total effective integration time (hours per unit solid angle of one square degree), in intensity and in polarization, achieved in each region after data flagging for each channel in the nominal mode (n) and in the combination of nominal plus raster scans (n+r), and final map sensitivities in Stokes $I$, $Q$ and $U$ maps calculated through a null test analysis (see text for details). The first digit of the channel identifier refers to the horn index and the next two digits indicate the frequency.} 
\begin{center}
\begin{tabular}{cccccccccc} 
	\hline\noalign{\smallskip}
	 Channel & \multicolumn{5}{c}{Integration times} && \multicolumn{3}{c}{Sensitivity}        \\
	               & \multicolumn{5}{c}{(h\,deg$^{-2}$)} && \multicolumn{3}{c}{($\mu$K\,deg$^{-1}$)}        \\
	               \noalign{\smallskip}
	 \cline{2-6}\cline{8-10}
	 \noalign{\smallskip}
                        & \multicolumn{2}{c}{$I$} &&   \multicolumn{2}{c}{$Q$,$U$} && $I$ & $Q$ & $U$\\
                        \cline{2-3}\cline{5-6}
	& n & n+r && n & n+r &&&\\
	\hline\noalign{\smallskip}
        \multicolumn{10}{c}{$\rho$\,Ophiuchi}                    \\\noalign{\smallskip}
        \noalign{\smallskip}
	217	&  0.15 &  2.0 && 0.14 &  1.4 && 38   & 16    & 17   \\ 
	219 	&  0.13 &  2.0 && 0.12 &  1.6 && 47   & 19    & 22   \\
	311     &  0.13 &  1.3 && 0.11 &  1.1 && 26   & 10    & 10   \\
	313	&  0.10 &  1.3 && 0.08 &  1.1 && 19   & 7     & 10   \\
	417	&  0.09 &  1.2 && 0.07 &  0.8 && 125  & 12    & 12   \\
	419	&  0.04 &  1.1 && 0.03 &  0.7 && 136  & 15    & 16   \\
	\hline\noalign{\smallskip}
        \multicolumn{10}{c}{Perseus}                        \\\noalign{\smallskip}
	217	&  0.20 &  4.2 && 0.11 &  1.54 && 22  & 6     & 6    \\
        219 	&  0.18 &  2.6 && 0.10 &  0.88 && 33  & 10    & 10   \\
	311     &  0.18 &  2.1 && 0.12 &  0.70 && 22  & 10    & 8    \\
	313	&  0.16 &  1.9 && 0.11 &  0.62 && 19  & 10    & 8    \\
	417	&  0.20 &  0.9 && 0.11 &  0.11 && 55  & 22    & 23   \\
	419	&  0.16 &  0.7 && 0.09 &  0.09 && 69  & 23    & 24   \\
 	\hline\noalign{\smallskip}
        \multicolumn{10}{c}{W43}                           \\\noalign{\smallskip}
	      217	  &  0.22 & 0.83 && 0.10 & 0.40 &&  46  & 12    & 13  \\
        219 	&  0.19 & 0.80 && 0.09 & 0.39 &&  61  & 15    & 24  \\
	    311     &  0.08 & 0.36 && 0.05 & 0.20 &&  56   & 46    & 42  \\
        313     &  0.09 & 0.38 && 0.06 & 0.21 &&  40   & 37    & 38  \\
	      417	  &  0.26 & 0.85 && 0.12 & 0.12 &&  52   & 22    & 22  \\
	    419	    &  0.22 & 0.81 && 0.10 & 0.10 &&  62   & 24    & 24  \\
        \hline
\end{tabular}
\end{center}
\label{tab: sensitivity}
\end{table}	

\subsection{Ancillary data}

\subsubsection{Low-frequency radio surveys}
Data in total intensity at frequencies below QUIJOTE-MFI are needed to model the free-free emission\footnote{As explained in section~\ref{subsec: Foreground modelling} the three regions that are studied in this paper are fully dominated by free-free emission, and do not show evidence of any synchrotron emission}. At these frequencies we used the surveys listed in Table~\ref{tab: surveys summary}: i) the full-sky ``Haslam'' map at 0.408\,GHz \citep{haslam1982}, ii) the ``Dwingeloo'' 0.820\,GHz map of the northern sky \citep{dwingeloo}, iii) the ``Reich'' map of the northern sky at 1.42\,GHz \citep{Reich_86}, iv) the S-PASS survey of the southern sky at 2.3\,GHz \citep{carretti2019} and v) the ``HartRAO'' map of the southern sky at 2.326\,GHz \citep{hartrao}. For the Haslam, Reich and HartRAO maps we used the public versions of \cite{plataniamaps}.  The data from the Dwingeloo survey have been extracted from the MPIfR's Survey Sampler\footnote{http://www.mpifr-bonn.mpg.de/survey.html} and projected into a {\tt HEALPix} pixelization. The S-PASS maps were downloaded from the LAMBDA database\footnote{\tt https://lambda.gsfc.nasa.gov}. As we do for all other surveys, these maps are convolved to a common angular resolution of $1^\circ$, except the Dwingeloo map whose native angular resolution is $1.2^\circ$. The slightly larger angular resolution of this map may have an impact on the derived results that is accounted for in the 10\% calibration uncertainty that is assigned to this map (see Table~\ref{tab: surveys summary}).

Except for the Haslam map all these surveys have a partial sky coverage. The Dwingeloo map does not cover the $\rho$\,Ophiuchi region, while neither the S-PASS nor the HartRAO surveys cover the Perseus region. For $\rho$\,Ophiuchi and W43 the flux densities of these two last surveys were averaged into one single measurement at 2.3\,GHz. For W43 we also used the C-BASS \citep{Jones_18} flux densities extracted by \cite{Irfan2015} appropriately rescaled in intensity, as explained in \cite{W44}. The calibration of both  the Reich and the HartRAO maps is referenced to the full-beam solid angle. To overcome this issue, and translate the calibration to the main-beam, we multiply the Reich map by 1.55 \citep{reich1988}. In the case of the HartRAO map, for $\rho$\,Ophiuchi and Perseus we have applied the standard factor of 1.45 derived by \cite{hartrao}, while in W43 we have applied a smaller factor of 1.2 to account for the fact that its angular size is larger than the telescope's beam (see related discussion in \citealt{W44}). Uncertainties on these factors are accounted for in the 10\% calibration uncertainties assigned to these maps (see Table~\ref{tab: surveys summary}). Other systematic effects that affect these maps are uncertainties related to the determination of zero levels, but our analyses are insensitive to this thanks to the subtraction of an average background level through our aperture photometry technique (see section~\ref{subsec: photometry}).  

\subsubsection{Microwave, mm and sub-mm surveys: WMAP, \textit{Planck} and DIRBE}

In the microwave regime we used data from WMAP and \textit{Planck}, which helps to better constrain the AME spectrum, and in the mm and sub-mm ranges we used, in addition to \textit{Planck}, data from COBE-DIRBE that allows us to model the spectrum of the thermal dust emission. 

The WMAP satellite produced full-sky maps, in intensity and polarization, at 23, 33, 41, 61 and 94\,GHz \citep{wmap}. In this analysis we use the version of WMAP 9-year maps smoothed to a resolution of $1^\circ$ that are available from the LAMBDA database\footnote{\tt http://lambda.gsfc.nasa.gov}. The \textit{Planck} mission \citep{cpp2018-1} produced full-sky maps at central frequencies of 28, 44, 70, 100, 143, 217, 353, 545 and 857\,GHz in total intensity, and in polarization in the seven lower-frequency bands. In intensity we use maps from the \textit{Planck} 2015 data release (PR2), including the Type 1 CO maps that are used to correct the 100, 217 and 353\,GHz intensity maps from the contamination introduced by the CO rotational transition lines $J=$\,1$\rightarrow$0, $J=$\,2$\rightarrow$1 and $J=$\,3$\rightarrow$2 respectively. In polarization we used the \textit{Planck} 2018 data release (PR3). This choice is motivated by the fact that in the LFI frequencies we have applied our own implementation of the leakage correction in polarization, for which we have used the projection maps that are available only for PR3 (see section~\ref{subsec:leakage_corr}). There is a negligible difference between using PR2 or PR3 to extract flux densities of compact sources in total intensity (see e.g. \citealt{AMEwidesurvey}). These \textit{Planck} maps have been downloaded from the \textit{Planck} Legacy Archive (PLA)\footnote{\tt https://pla.esac.esa.int}.    

The spectral coverage of DIRBE, an infrared instrument onboard the COBE satellite, spans from 1.25 to 240\,$\mu$m \citep{cobe-dirbe}. We used maps at 240 $\mu$m (1249\,GHz), 140 $\mu$m (2141\,GHz) and 100 $\mu$m (2997\,GHz), that are the three frequencies dominated by the population of big grains that can be modelled with a single modified blackbody spectrum. We have used the zodiacal-light subtracted mission average (ZSMA) maps regridded into the {\tt HEALPix} format. 

Table~\ref{tab: surveys summary} lists the calibration uncertainties ascribed to each of these surveys that have been used in the subsequent analyses. They are the same used in previous recent works by the QUIJOTE collaboration (see e.g. \cite{AMEwidesurvey} and references therein).

\section{Methodology} \label{sec: METHODOLOGY}

\subsection{Flux-density estimation through aperture photometry} \label{subsec: photometry}

Intensity and polarization flux densities are calculated through a standard aperture photometry method applied on the $1^\circ$-smoothed maps of each region. This is a well-known and widely used technique \citep{caraballo2011,planck2011XXnewlight,Perseus, W44, Taurus,snrwidesurvey} consisting in integrating temperatures of all pixels within a given aperture, and subtracting a background level estimated through the median of all pixels in an external ring. The flux density is then given by
\begin{equation} \label{eq:flux_i}
   S_{\nu} = a(\nu) \left( \frac{\sum_{i=1}^{n_{1}}{T_{i}}}{\rm n_{1}}- \tilde{T} \right) ,
\end{equation}
\\ 
where 
\begin{equation}
   a(\nu) = \frac{x^2 e^x}{(e^x - 1)^2}\left(\frac{2 k_{B}\nu^{2}}{c^{2}}\right){\rm n_{1}}\Omega_{\rm pix} ,
\end{equation}
\\ 
is the conversion factor between thermodynamic differential temperature (K$_{\rm CMB}$ units) and flux-density (units of 10$^{26}$\,Jy), $T_i$ is the thermodynamic temperature of pixel $i$ inside the aperture, $n_{1}$ is the number of pixels in the aperture, $\tilde{T}$ is the median temperature of all pixels in the background region, $\Omega_{\rm pix}$ is the solid angle corresponding to one pixel and $x = h \nu / (k_{\rm B} T_{\rm CMB})$ is the dimensionless frequency.

We have considered two different methods to estimate the error of $S_{\nu}$. The first one is based on the analytical propagation of pixel errors through the equation
\begin{equation}
   \sigma_{\rm stat}({S_\nu})  = a(\nu)\, \sigma(T) \left[ \frac{1} {n_{1}}  + \frac{\pi}{2} \frac{1}{n_{2}} \right]^{1/2} ~~,
   \label{eq:sigma_snu}
\end{equation}
where $\sigma(T)$ is the error of the temperature value of each pixel that is considered uniform and is derived from the pixel-to-pixel standard deviation calculated in the background ring, and $n_2$ is the total number of pixels in the background annulus. This equation assumes perfectly uncorrelated noise between pixels. As explained in section \ref{sec:maps}, in general the noise is spatially correlated due to the presence of $1/f$ residuals. In addition background fluctuations on scales larger than the pixel size also introduce correlated noise. The noise correlation function could be introduced in equation~\ref{eq:sigma_snu}, but its determination is not trivial. Alternatively, as a second method that accounts jointly for both contributions ($1/f$ and white noise), we derive flux densities in ten apertures located around the source, using the same aperture and external annulus radii, and derive $\sigma_{\rm stat}({S_\nu})$ through the scatter of these estimates. We have applied this method to estimate uncertainties in the polarization flux density estimates. In total intensity we have used this same method in $\rho$\,Ophiuchi and Perseus. In W43 we found out that uncertainties using equation~\ref{eq:sigma_snu} lead to a global fit with reduced $\chi^2$ close to one so in this case we decided to stick to this method. Details related with the calculation of the flux-density errors of each region will be explained in the corresponding sections.

The calibration uncertainty of each survey is combined with the statistical error to derive a final global error as 
\begin{equation}
\sigma(S_\nu)=\sqrt{\sigma_{\rm stat}(S_\nu)^2+(\delta\cdot S_\nu)^2}~~, 
\label{eq:flux_error}
\end{equation}
where $\delta$ is the calibration fractional error (quoted in Table~\ref{tab: surveys summary} in percent units).

Central coordinates and sizes of the circular aperture and of the inner and outer circles of the background ring are given in Table~\ref{tab: AP_configurations}. In general we have opted to choose the same values as in previous studies of the same regions to allow for a more reliable comparison with previous results. For $\rho$\,Ophiuchi we have used the same parameters as in \cite{planck2011XXnewlight} and in \cite{Dickinson2011}. In this case, to obtain a more realistic background estimate we have removed the emission from the nearby HII region, which is brighter at QUIJOTE-MFI frequencies, by masking all pixels lying in a circle of radius $0.4^\circ$ around the position $(l,b) = (351.5^\circ$, $17.05^\circ$). In the case of Perseus we used the same configuration as in the intensity analysis of \cite{Perseus}, while for W43 we used that of \cite{W44}. Intensity flux densities for the three regions are shown in Table~\ref{tab:fluxes_i}\footnote{In this Table, the 2.3\,GHz value in the case of $\rho$\,Ophiuchi is the weighted average of the flux densities derived from S-PASS and HartRAO. In the case of W43 the value comes from HartRAO, as the S-PASS map does not cover this region.}, while flux densities calculated on $Q$ and $U$ maps are shown in Tables~\ref{tab:rhoophw_pol_constraints}, \ref{tab:perseus_pol_constraints} and \ref{tab:w43_pol_constraints}.

\begin{table*}
\caption{Intensity flux densities for $\rho$\,Ophiuchi, Perseus and W43, obtained through aperture photometry on maps degraded to a common angular resolution of $1^\circ$.}
\begin{center}
\begin{tabular}{cccccc}
	\hline
	Freq.		& $\rho$\,Ophiuchi  	                   & Perseus                              & W43                            \\
	(GHz)		& (Jy) 		               	           & (Jy)				                  & (Jy)                           \\		
	\hline
	0.408 		& -7.2 $\pm$ 7.6			           & 9.5 $\pm$ 8.1                        & 496 $\pm$ 54                       \\
        0.82 		& -                                    & 10.1 $\pm$ 4.7                       & 444 $\pm$ 48                       \\
	1.42 		& 0.0 $\pm$ 6.9				           & 9.7 $\pm$ 3.2                        & 391 $\pm$ 43                       \\
	2.3 		& 0.4 $\pm$ 2.3 			           & -       	                          & 471 $\pm$ 43                       \\
        4.76        & -                                    & -                                    & 400 $\pm$ 52                       \\
	11.1		& 7.9 $\pm$ 1.5				           & 14.6 $\pm$ 2.4                       & 437 $\pm$ 26                       \\
	12.9 		& 10.4 $\pm$ 1.7			           & 18.0 $\pm$ 2.2                       & 515 $\pm$ 43                       \\
	16.8		& 16.0 $\pm$ 1.9			           & 28.5 $\pm$	3.4                       & 519 $\pm$ 30                       \\
	18.8		& 20.7 $\pm$ 4.5		               & 34.2 $\pm$ 3.6                       & 535 $\pm$ 30                       \\
	22.8		& 27.0 $\pm$ 2.4			           &  37.8 $\pm$ 2.5 	                  & 525 $\pm$ 21                       \\
	28.4		& 30.3 $\pm$ 2.6			           &  38.2 $\pm$ 2.5                      & 533 $\pm$ 22                       \\
	33.0 		& 30.4 $\pm$ 2.7			           &  36.1 $\pm$ 2.7                      & 502 $\pm$ 21                       \\
	40.6		& 27.5 $\pm$ 2.8			           &  32.1 $\pm$ 3.7                      & 472 $\pm$ 19	                   \\
	44.1		& 26.4 $\pm$ 3.1			           & 30.8 $\pm$ 4.4                       & 457 $\pm$ 18                       \\
	60.5 		& 26.6 $\pm$ 4.4			           & 30.5 $\pm$ 7.5                       & 418 $\pm$ 17                       \\
	70.4 		& 32.1 $\pm$ 5.6			           &  36.8 $\pm$ 10.0                     & 438 $\pm$ 18                       \\
	93.5		& 64.5 $\pm$ 8.4			           &  69.0 $\pm$ 15.0                     & 564 $\pm$ 23                       \\
	100 		& 80 $\pm$ 9				           &  80 $\pm$ 18                         & 627 $\pm$ 26                       \\
	143 		& 227 $\pm$ 17				           &  205 $\pm$ 36                        & (1.35 $\pm$ 0.06) $\times 10^{3}$  \\
	217 		& 917 $\pm$ 46				           &  786 $\pm$ 99                        & (4.86 $\pm$ 0.23) $\times 10^{3}$   \\
	353			& (4.31 $\pm$ 0.21) $\times 10^{3}$    &  (3.68 $\pm$ 0.43) $\times 10^{3}$   & (2.41 $\pm$ 0.11) $\times 10^{4}$    \\
	545 		& (1.55 $\pm$ 0.11)	$\times 10^{4}$	   &  (1.25 $\pm$ 0.16)	$\times 10^{4}$	  & (9.01 $\pm$ 0.58) $\times 10^{4}$    \\
	857			& (5.33 $\pm$ 0.40)	$\times 10^{4}$	   &  (4.04 $\pm$ 0.49) $\times 10^{4}$	  & (3.59 $\pm$ 0.25) $\times 10^{5}$   \\
	1249		& (1.21 $\pm$ 0.15) $\times 10^{4}$	   &  (8.30 $\pm$ 0.90) $\times 10^{4}$	  & (9.28 $\pm$ 0.11) $\times 10^{5}$    \\
	2141 		& (2.27 $\pm$ 0.25) $\times 10^{5}$    &  (1.14 $\pm$ 0.12) $\times 10^{5}$   & (1.87 $\pm$ 0.22) $\times 10^{6}$    \\
	2997		& (1.45 $\pm$ 0.20)	$\times 10^{5}$	   &  (5.64 $\pm$ 0.75)	$\times 10^{4}$   & (1.06 $\pm$ 0.16) $\times 10^{6}$   \\
	\hline
\end{tabular}
\end{center}
\label{tab:fluxes_i}
\end{table*}

\subsection{SED modelling in total intensity}
\label{subsec: Foreground modelling}
We modelled four different components in our frequency range, between 0.4 and 3000\,GHz: free-free, anomalous microwave emission (AME), thermal dust and CMB anisotropies. The low-frequency spectra of the three molecular cloud complexes studied in this paper are fully dominated by free-free emission, and therefore the synchrotron emission is not considered in the fits. The physical models used for each of these components are briefly explained in the following subsection.

\subsubsection{Sky model} 
\label{subsubsec: Sky model}

\paragraph{Free-free emission.}
\hspace{1cm}~\\
\label{subsubsec: free-free}
\noindent Taking into account that $T_{\rm e}(1-e^{-\tau_{\rm ff}})$ is the brightness temperature of the free-free emission for a medium with optical depth $\tau_{\rm ff}$ and electron temperature $T_{\rm e}$, the corresponding flux density can be calculated as: 
\\
\begin{equation}
S_{\nu}^{\rm ff}({\rm EM}) = \frac {2k_{\rm B} \nu^{2}} {c^{2}} \,\Omega\, T_{\rm e} (1-e^{-\tau_{\rm ff}})~~.
\label{eq:sff}
\end{equation}
Here we have considered the equations derived by \cite{draine2011} for the optical depth,
\begin{equation}
\tau_{\rm ff} = 5.468 \cdot 10^{-2} \cdot {\rm EM} \cdot (T_{\rm e})^{-3/2} \cdot \left( \frac {\nu}{\rm GHz} \right)^{-2} \cdot g_{\rm ff} (\nu)~~,
\label{eq:tauff}
\end{equation}
and for the Gaunt factor,
\begin{equation}
g_{\rm ff} (\nu) = {\rm ln}\left({\rm exp} \left( 5.960 - \frac {\sqrt{3}}{\pi} \cdot {\rm ln} \left[\left(\frac {\nu}{\rm GHz}\right) \left(\frac{T_{\rm e}}{10^{4} {\rm K}}\right)^{-3/2} \right] \right) + e\right).
\label{eq:gaunt_factor}
\end{equation}
For the electron temperature we have used $T_{\rm e}=8000$ K (same value as \citealt{planck2011XXnewlight} and  \citealt{Perseus}) for $\rho$ Ophiuchi and Perseus and $T_{\rm e}=6038$ K for W43; this last value is the same used in \citep{W44} and is extracted from a template of the free-free emission at 1.4 GHz produced by \citep{Alves2012} using radio recombination line data from the HI Parkes All-Sky Survey (HIPASS). The only remaining free parameter associated with the free-free component is the emission measure EM (units of pc $\cdot$ cm$^{-6}$).

\paragraph{Thermal Dust.}
\hspace{1cm}~\\
\label{subsubsec: thermal dust}
\noindent Following the common practice in the field (see e.g. \citealt{cpp2013-11}) the thermal dust emission is modelled as a single-component modified blackbody (MBB) curve, $\nu^{\beta_{\rm d}} B_\nu(\nu,T_{\rm d})$, which we normalized using the optical depth at 250 $\mu$m (1.2 THz), $\tau_{250}$:
\begin{equation}
S_{\nu}^{\rm dust}(\beta_{\rm d}, T_{\rm d}, \tau_{250}) = \frac {2h\nu^3}{c^2} \frac {1}{e^{{h}\nu/{k_{\rm B}T_{\rm d}}}-1} \tau_{250} \left(\frac {\nu} {1.2\ {\rm THz}} \right)^{\beta_{{\rm d}}} \Omega\ ,
\end{equation}
\\
where the dust temperature $T_{\rm d}$ and the emissivity index $\beta_{\rm d}$, together with $\tau_{250}$, are the three free parameters.

\paragraph{AME} 
\hspace{1cm}~\\
\label{subsubsec: AME}
\noindent Here, we have modelled the AME through a phenomenological model consisting in a parabola in the log(S) - log($\nu$) plane  \citep{stevenson2014}:
\begin{equation}
\begin{split}
S_{\nu}^{\rm AME}&(A_{\rm AME}, \nu_{\rm AME}, W_{\rm AME}) = \\ & {A_{\rm AME}} \cdot
{\rm exp} \left[ -\frac {1}{2 {W_{\rm AME}}^2} {\rm ln}^2 \left(\frac{\nu}{\nu_{\rm AME}}\right) \right] ,  
\end{split}
\end{equation}
where $A_{\rm AME}$ is the maximum flux density, $\nu_{\rm AME}$ the correspondent frequency for that maximum and $W_{\rm AME}$ the width of the parabola on the log-log plane. This phenomenological model reproduces with high fidelity the spinning dust models and, thanks to its simplicity and due to the difficulty of jointly fitting the large number of parameters of those models, is frequently used by other recent studies \citep{LambdaOrionis,AMEwidesurvey,ameplanewidesurvey}.

\paragraph{CMB}
\hspace{1cm}~\\
\label{subsubsec: CMB}
Although the CMB monopole (constant) term is cancelled in the background subtraction in our photometry method (see Sect.\ref{subsec: photometry}), CMB fluctuations could still have a contribution in the angular scale of the aperture. They are then modelled as
\begin{equation}
   \Delta S_{\nu}^{\rm CMB}({\Delta T}_{\rm CMB}) =  \frac{x^2 e^x}{(e^x - 1)^2} \left(\frac{2 k_{B}\nu^{2}}{c^{2}}\right) {\Delta T}_{\rm CMB} \Omega ~~,
\end{equation}
where the fitted parameter is the amplitude $\Delta T_{\rm CMB}$.

\begin{table*}
\caption{AME (between frequencies 11.1 and 44.1\,GHz) and thermal dust (frequencies 60.5 to 353\,GHz) polarization constraints on $\rho$\,Ophiuchi. The second column shows residual AME ($S_\nu^{\rm AME}$) or residual thermal dust ($S_{\nu}^{\rm dust}$ ) flux densities. The next columns list flux densities in $Q$ and $U$, debiased polarized flux densities and debiased AME or thermal dust polarization fractions. For both the polarized flux density and the polarization fraction the reported uncertainties and upper limits are referred respectively to the 68\% and 95\% confidence levels. We also show the polarization constraint on $\Pi_{\rm AME}$ derived from the stacked map (see Sect.~\ref{sec:stacking} for details).}
\label{tab:rhoophw_pol_constraints}
\begin{center}
\begin{tabular}{cccccc}
	\hline
	Freq.		& $S_\nu^{\rm AME}$  or $S_{\nu}^{\rm dust}$  	& $Q$ 				    & $U$ 				    & $P_{\rm db}$ 		& $\Pi_{\rm AME}$  or   $\Pi_{\rm dust}$  \\
	(GHz)		& (Jy) 				& (Jy)			        & (Jy)			        & (Jy)			    & (\%)		         \\ 
	\hline
	&\multicolumn{5}{c}{AME}\smallskip\\
	11.1		& 6.4 $\pm$ 1.5 	& 0.08 $\pm$ 0.19       & 0.45 $\pm$ 0.21	    & 0.40$^{+0.22}_{-0.20}$	& < 12.8            \\
	12.9        & 8.8 $\pm$ 1.7     & 0.18 $\pm$ 0.14       & 0.37 $\pm$ 0.20       & 0.37$^{+0.18}_{-0.17}$    & < 7.9             \\
	16.8	    & 14.2 $\pm$ 1.9	& $-0.08$ $\pm$ 0.18	& 0.38 $\pm$ 0.28	    & < 0.71	                & < 5.1		     \\
	18.8		& 18.8 $\pm$ 4.5	& 0.32 $\pm$ 0.46		& 0.46 $\pm$ 0.62	    & < 1.33          	        & < 7.4	         \\
	22.8		& 24.6 $\pm$ 2.5 	& 0.08 $\pm$ 0.13		& 0.28 $\pm$ 0.09	    & 0.26$^{+0.12}_{-0.11}$	& < 1.9	         \\
	28.4		& 27.1 $\pm$ 2.8	& 0.001 $\pm$ 0.140		& 0.10 $\pm$ 0.11	    & < 0.29          	        & < 1.0	         \\
	33.0 		& 26.3 $\pm$ 3.0	& $-0.23$ $\pm$ 0.14    & 0.02 $\pm$ 0.14	    & < 0.43	                & < 1.7	         \\
	40.6		& 21.2 $\pm$ 3.4	& 0.04 $\pm$ 0.24	   	& $-0.12$ $\pm$ 0.23	& < 0.50    	            & < 2.4             \\
	44.1		& 18.8 $\pm$ 3.8	& 0.13 $\pm$ 0.15	    & 0.14 $\pm$ 0.22	    & < 0.46 	                & < 2.5	         \\
    22.8 (stack)  &  27.2 $\pm$ 1.5      &  $-0.030$ $\pm$ 0.076  &  0.033 $\pm$ 0.067  &          < 0.16           & < 0.58             \\
	\hline
	&\multicolumn{5}{c}{Thermal dust}\smallskip\\
	60.5		& 9.9 $\pm$ 1.6     & 0.36 $\pm$ 0.37	    & 0.68 $\pm$ 0.41 		& 0.64$^{+0.45}_{-0.35}$	& < 14.0	        \\
	70.4		& 16.9 $\pm$ 2.8    & 0.47 $\pm$ 0.22	    & $-0.19$ $\pm$ 0.27	& 0.42$^{+0.28}_{-0.23}$    & 2.5$^{+1.7}_{-1.5}$	 \\
	93.5		& 45.3 $\pm$ 7.8    & 2.02 $\pm$ 0.89	    & 0.45 $\pm$ 1.13		& 1.75$^{+1.15}_{-0.96}$	& 4.1$^{+2.8}_{-2.1}$   \\
	100		    & 58.5 $\pm$ 10.1	& 0.90 $\pm$ 0.24	    & $-2.34$ $\pm$ 0.33	& 2.51 $\pm$ 0.26		    & 4.4 $\pm$ 0.9	    \\
	143		    & 204 $\pm$ 37      & 3.12 $\pm$ 0.55	    & $-1.49$ $\pm$ 0.79	& 3.39 $\pm$ 0.67           & 1.7$^{+0.4}_{-0.5}$	    \\
	217		    & 851 $\pm$ 169		& 9.91 $\pm$ 1.85 	    & $-6.15$ $\pm$ 2.55	& 11.5 $\pm$ 2.2		    & 1.4 $\pm$ 0.4     \\
	353		    & 4234 $\pm$ 986	& 42.7 $\pm$ 8.8	    & $-33.1$ $\pm$ 14.0	& 52.9 $\pm$ 11.3		    & 1.3 $\pm$ 0.4     \\
	\hline
\end{tabular}
\end{center}
\end{table*}

\begin{table*}[h!]
\caption{Same as in Table~\ref{tab:rhoophw_pol_constraints} but for Perseus.}
\label{tab:perseus_pol_constraints}
\begin{center}
\begin{tabular}{cccccc}
	\hline
	Freq.		& $S_\nu^{\rm AME}$  or $S_{\nu}^{\rm dust}$  	& $Q$ 				    & $U$ 				    & $P_{\rm db}$ 		& $\Pi_{\rm AME}$  or   $\Pi_{\rm dust}$  \\
	(GHz)		& (Jy) 				    & (Jy)				    & (Jy)				    & (Jy) 			    & (\%) 			    \\	
\hline
	&\multicolumn{5}{c}{AME}\smallskip\\
	11.1			&   6.7 $\pm$ 2.4 		&$-0.24 \pm 0.09$		& $-0.29 \pm 0.14$		&  < 0.56			& < 10.7 	  		\\
	12.9 			&  10.0 $\pm$ 2.2       &$-0.05 \pm 0.26$		& $-0.08 \pm 0.21$		&  < 0.48			& < 4.8 		    \\	
	16.8			&  20.3 $\pm$ 3.4		&$ 0.21 \pm 0.31$		& $ 0.04 \pm 0.32$		&  < 0.69			& < 3.5  		    \\
	18.8			&  25.8 $\pm$ 3.7		&$-0.10 \pm 0.56$		& $ 0.13 \pm 0.33$		&  < 0.89			& < 3.5 			\\
	22.8			&  28.9 $\pm$ 2.7 	    &$-0.03 \pm 0.15$	    & $-0.08 \pm 0.14$    	&  < 0.31	        & < 1.1  			\\
	28.4			&  28.2 $\pm$ 2.9		&$-0.11 \pm 0.16$	    & $-0.08 \pm 0.13$	    &  < 0.35	        & < 1.2  			\\
	33.0 			&  25.0 $\pm$ 3.4		&$ 0.14 \pm 0.18$   	& $-0.05 \pm 0.24$  	&  < 0.47           & < 1.9 			\\
	40.6			&  18.3 $\pm$ 4.8		&$-0.25 \pm 0.40$	    & $-0.34 \pm 0.27$    	&  < 0.89           & < 5.2 			\\
	44.1			&  15.5 $\pm$ 5.7		&$ 0.07 \pm 0.45$	    & $-0.62 \pm 0.38$    	&  < 1.21           & < 9.0  			\\
    22.8 (stack)           &        28.14 $\pm$ 1.77          & 0.001 $\pm$ 0.086   &  -0.310 $\pm$ 0.101   &      < 0.45             & < 1.64            \\
        \hline
        	&\multicolumn{5}{c}{Thermal dust}\smallskip\\
	60.5    & 7.3 $\pm$ 2.3             & 1.14 $\pm$ 0.90	        & $-0.88 \pm 0.7$ 		& 1.04$^{+0.98}_{-0.58}$	 	& < 40.3              \\
	70.4	& 12.7 $\pm$ 4.1            & 0.49 $\pm$ 0.62 	        & $-0.02 \pm 0.7$		& < 1.51           		        & < 12.9              \\
	93.5	& 35.5 $\pm$ 11.7           & 0.76 $\pm$ 1.55	        & $-2.33 \pm 2.4$		& < 5.24            	        & < 16.6              \\
	100		& 45.3 $\pm$ 15.1	        & 1.10 $\pm$ 0.47	        & $-0.40 \pm 0.7$		& 0.98$^{+0.66}_{-0.54}$		& 2.3$^{+1.7}_{-1.4}$	       \\
	143		& 162 $\pm$ 57		        & 7.02 $\pm$ 1.07	        & $-6.25 \pm 1.9$		& 9.40 $\pm$ 1.58               & 6.3$^{+2.1}_{-2.2}$	     \\
	217		& 693 $\pm$ 271	            & 31.1 $\pm$ 3.8  	        & $-29.9 \pm 8.0$	    & 43.1 $\pm$ 6.4		        & 7.3$^{+2.6}_{-2.3}$	     \\
	353		& 3494 $\pm$ 1666	        & 172 $\pm$ 17        	    & $-125  \pm 34 $       & 213 $\pm$ 29		            & 7.3 $\pm$ 2.8     \\
	\hline	
\end{tabular}
\end{center}
\end{table*}

\begin{table*}
\caption{Same as in Table~\ref{tab:rhoophw_pol_constraints} but for W43.}
\label{tab:w43_pol_constraints}
\begin{center}
\begin{tabular}{cccccc}
	\hline
	Freq.		& $S_\nu^{\rm AME}$  or $S_{\nu}^{\rm dust}$  	& $Q$ 				    & $U$ 				    & $P_{\rm db}$ 		& $\Pi_{\rm AME}$  or   $\Pi_{\rm dust}$  \\
	(GHz)		    & (Jy) 				& (Jy)			        & (Jy)			        & (Jy)			        & (\%)		         \\		
	\hline
		&\multicolumn{5}{c}{AME}\smallskip\\
	16.8		    & 187 $\pm$ 30		&$-0.16 \pm 0.30$  	    & $-1.02 \pm 0.34$      &  0.97$^{+0.34}_{-0.33}$	      & <	0.85             \\
	18.8		    & 207 $\pm$ 30		&$ 0.17 \pm 0.76$  	    & $1.69  \pm 0.76$     	&  1.49$^{+0.85}_{-0.77}$	 	  & <	1.42             \\
	22.8			& 205 $\pm$ 21 		&$ 0.56 \pm 0.18$   	& $-0.28 \pm 0.11$      &  0.61 $\pm$ 0.14		          & <	0.43             \\
	28.4			& 220 $\pm$ 22		&$ 0.26 \pm 0.34$ 		& $0.41  \pm 0.20$		&  0.38$^{+0.31}_{-0.21}$	 	  & < 0.40             \\
	33.0 			& 193 $\pm$ 21		&$ 0.20 \pm 0.22$    	& $-0.22 \pm 0.13$	    &  0.22$^{+0.20}_{-0.12}$		  & <	0.29             \\
	40.6			& 164 $\pm$ 19		&$ 0.10 \pm 0.11$ 	   	& $-0.30 \pm 0.18$ 	    &  0.28$^{+0.15}_{-0.14}$	 	  & <	0.33             \\
	44.1			& 148 $\pm$ 19 		&$ 0.13 \pm 0.40$	    & $0.56  \pm 0.31$ 		&  < 1.08 		                  & <	0.73             \\
    60.5            & 86 $\pm$ 20       &$ 0.20 \pm 0.37$ 		& $0.22  \pm 0.45$ 		&  < 0.92                         & < 1.11             \\
    70.4            & 74 $\pm$ 25       &$-0.46 \pm 0.89$       & $1.34  \pm 0.87$       &  < 2.68                         & < 4.22             \\
    22.8 (stack)           & 184 $\pm$ 18    &  0.087 $\pm$ 0.090   &      0.425 $\pm$ 0.051   &   < 0.54                        & < 0.31             \\
      \hline
        	&\multicolumn{5}{c}{Thermal dust}\smallskip\\	
	100		    & 306 $\pm$ 52 		     & 2.50 $\pm$ 0.25  	 & $-2.96 \pm 0.27$     & 3.87 $\pm$ 0.26 	     & 1.27 $\pm$ 0.23     \\
	143			& 1092 $\pm$ 194         & 8.46 $\pm$ 0.56   	 & $ 1.19 \pm 0.54$     & 8.54 $\pm$ 0.54		 & 0.78 $\pm$ 0.15     \\
	217			& 4693 $\pm$ 902         & 36 $\pm$ 3 		     & $   15 \pm 3	  $     & 39 $\pm$ 3 		     & 0.86 $\pm$ 0.17     \\
	353 		& 24235 $\pm$ 5434 	     & 120 $\pm$ 14    	     & $   39 \pm 12  $     & 127 $\pm$ 13		     & 0.54 $\pm$ 0.13     \\
	\hline
\end{tabular}
\end{center}
\end{table*}	

\subsubsection{Model selection}
\begin{figure*}
\begin{center}
\includegraphics[width=17cm]{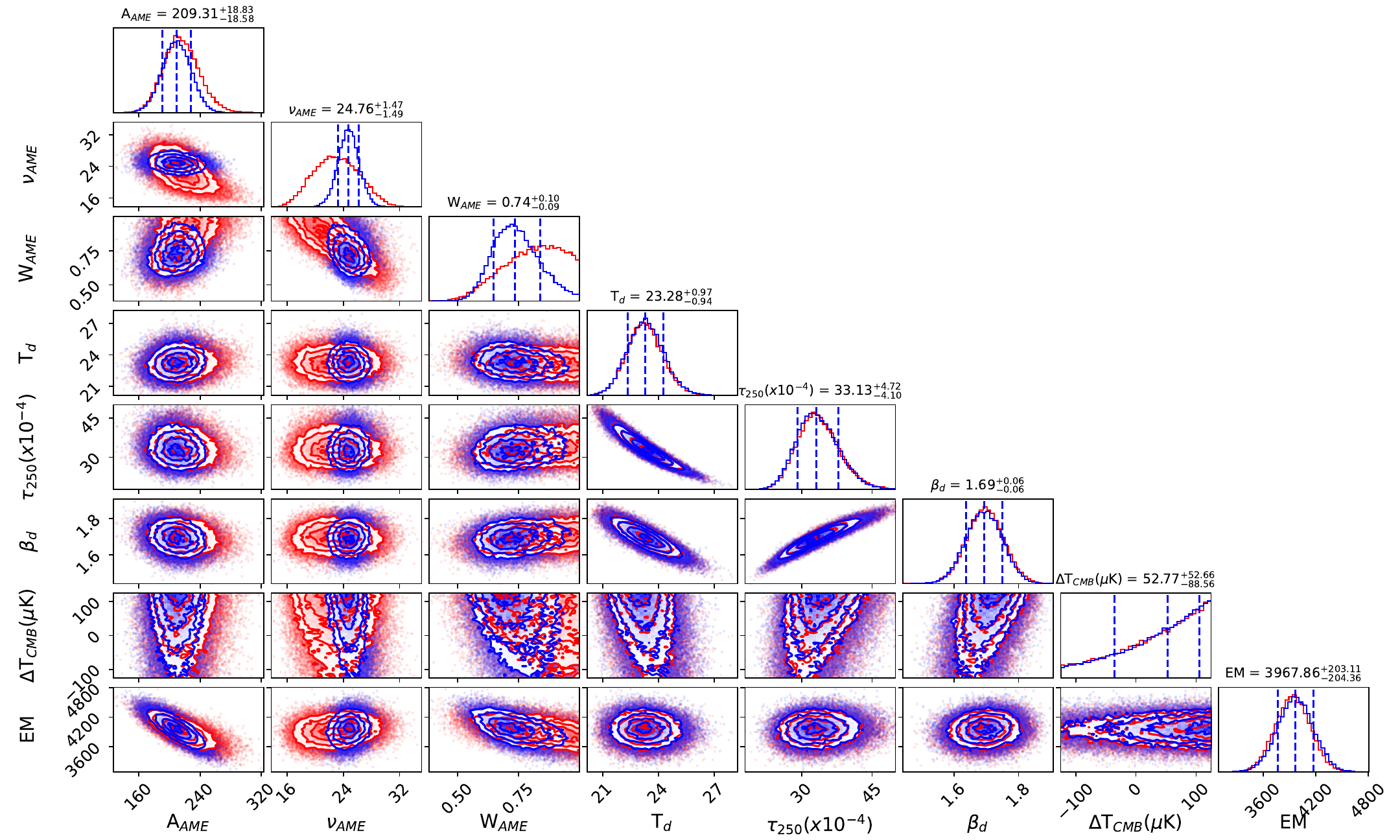}
\caption{Example of corner plot of the two-dimension parameter space explored by the MCMC implemented in the {\sc emcee} package, corresponding to the W43 molecular complex. Blue and red contours correspond respectively to the fits with and without QUIJOTE data (see derived best-fit parameters in Table~\ref{tab:model_parameters}). Also shown are one-dimension marginalised posterior distributions from which the best-fit parameters and uncertainties are determined.}
\label{fig:w43_corner_plot}
\end{center}
\end{figure*}

As described in the previous subsections our model consists of 8 free parameters: EM for the free-free emission, A$_{\rm AME}$, $\nu_{\rm AME}$ and W$_{\rm AME}$ for the AME, $\tau_{250}$, $\beta_{\rm d}$, and $T_{\rm d}$ for the thermal dust emission and $\Delta T_{\rm CMB}$ for the CMB anisotropies. To sample the parameter posterior distributions we used the MCMC sampler from the {\sc emcee} package \citep{emcee}. Table~\ref{tab: Parameter priors} shows the top-hat priors that we have placed on each parameter. The prior on $\Delta T_{\rm CMB}$ is used only in the case of W43, and indeed in this case the Markov chain tends to adopt values in the positive border of this interval (see Figure~\ref{fig:w43_corner_plot}). Due to its small contribution relative to other components $\Delta T_{\rm CMB}$ is generally not well constrained, and as it was discussed in \cite{ameplanewidesurvey}, imperfections of the MBB model in the range $\sim 100-600$\,GHz could in some cases be absorbed by this component.
In the case of W43 we used a more stringent prior on the emission measure, 1000 < EM < 1500\,pc$\cdot$cm$^{-6}$, which is driven by the information based on the radio recombination line data of \cite{Alves2012} (see related discussion in \citealt{W44}). The final best-fit parameters are determined from the median values of the parameter posteriors, while their uncertainties are derived from the half difference of the 16 and 84 percentiles. In those cases where the distributions are quite asymmetric we have reported two different values for the negative and positive uncertainties. In Figure~\ref{fig:w43_corner_plot} we represent the probability density functions, in two and one dimensions, and best-fit parameters and their uncertainties, for the best-fit model of W43.

\begin{table}
\caption{Priors on model parameters used in the fitting procedure.}
\begin{center}
\begin{tabular}{c}
	\hline
	Parameter priors                         \\
	\hline
        EM > 0                                   \\
        $A_{\rm AME}$ > 0                        \\
	10.0 GHz < $\nu_{\rm AME}$ < 60.0 GHz    \\
	0.2 < $W_{AME}$ < 1.0   		     \\
	10 K < $T_{\rm d}$ < 40 K    		     \\
	0.0005 < $\tau_{250}$ < 0.005            \\
        1 < $\beta_{\rm d}$ < 3       		     \\
        $-125\,\mu$K < $\Delta T_{\rm CMB}$ < $125\,\mu$K     \\
        \hline
\end{tabular}
\end{center}
\label{tab: Parameter priors}
\end{table} 

\subsection{Colour-correction} \label{subsec:colour-correction}
We have applied colour corrections for all surveys except for the low-frequency ones (0.408 to 2.326\,GHz) where they are assumed to be unnecessary thanks to their narrower bandpasses (typically $\Delta\nu/\nu < 2\%$). Each flux density is multiplied by a colour-correction coefficient derived using the {\sc fastcc} code \citep{fastcc}. For frequencies below and above 100\,GHz we used two different approaches as described in section~3.3.2 of \cite{ameplanewidesurvey}. Briefly, for $\nu<100$\,GHz we assumed a power-law model and the colour-correction coefficient was calculated from the fitted spectral index at each frequency, while for $\nu>100$\,GHz the $\beta_{\rm d}$ and $T_{\rm d}$ fitted parameters of the MBB law are used to interpolate on a previously-computed 2D grid. Colour corrections depend on the fitted model so the process is applied iteratively until convergence is reached. Colour corrections are typically $\lesssim 2\%$ for QUIJOTE, WMAP and \textit{Planck}-LFI, and $\lesssim 10\%$ for \textit{Planck}-HFI and DIRBE, which have considerably larger bandwidths.

\subsection{Polarization analyses}
Flux densities in polarization were calculated for frequencies between 11\,GHz and 353\,GHz. In this section we describe specific tools that are applied to the analysis of polarization data.

\subsubsection{Noise debiasing of the polarized intensity}\label{sec:noise_debiasing}
Due to the polarized intensity $P=\sqrt{Q^{2}+U^{2}}$ being a positive-defined quantity, noise in the measurement of $Q$ and $U$ lead to a positive bias on the measured values of $P$ and of $\Pi=P/I$, which is more pronounced in the low signal-to-noise regime as it is our case. In this case knowledge of the full probability function of $P$ (which is no longer Gaussian even if errors of $Q$ and $U$ are Gaussian distributed) is needed in order to reliably determine the most-likely values and confidence intervals of $P$ and $\Pi$. We follow the same prescription that was described and applied in \cite{JARM2012} and previous QUIJOTE papers \citep[e.g.][]{Perseus,W44}. Specifically, to debias $P$ we follow a Bayesian approach consisting in integrating the analytical posterior probability density function (PDF) given in \cite{Vaillancourt2006}. For $\Pi$ we also integrate its PDF that, in this case, is evaluated through a Monte Carlo approach. In both cases we report best-fit values and 68\,\% errors determined from these PDFs when the signal-to-noise ratio of the measured quantity is larger than $\sqrt{2}$. Otherwise we will quote upper limits at the 95\% confidence level.

\subsubsection{Correction of intensity-to-polarization leakage in \textit{Planck} LFI}
\label{subsec:leakage_corr}
One of the most important systematic effects in polarization of \textit{Planck}-LFI is intensity-to-polarization leakage caused by the bandpass mismatch of the two orthogonally-polarized arms of the same radiometer (see e.g. \citealt{cpp2015-3}). Correction of this spurious signal requires knowledge of i) the spectrum of the emission in intensity, ii) the bandpasses of the two arms of the radiometer and iii) the scanning directions of each pixel to transform between sky and local coordinates. The way this correction is implemented is described in sections 11.1 to 11.4 of \citealt{cpp2015-2}. The corrected Stokes parameters are given by equation C.1 of \cite{cpp2015-26}:
\begin{equation}
\begin{pmatrix} Q \\ U \\ \end{pmatrix}_{\rm corr} = \begin{pmatrix} Q \\ U \\ \end{pmatrix}-\begin{pmatrix} P_Q \\ P_U \\ \end{pmatrix} (\alpha - \alpha_{\rm CMB})I~~,
\label{eq:qu_corr}
\end{equation}
where $Q_{\rm corr}$ and $U_{\rm corr}$ are the corrected maps, $Q$ and $U$ are the raw maps, $P_Q$ and $P_U$ are the leakage projection maps (see section 11.4 of \cite{cpp2015-2}), $\alpha$ is the spectral index of the sky emission (in flux-density units) in the considered frequency band and $\alpha_{\rm CMB}$ is the spectral index of the CMB (1.96, 1.90 and 1.75 at 28.4, 44.1 and 70.4\,GHz respectively). For PR2 and PR3 the leakage-correction maps at an angular resolution of 1$^\circ$ and $N_{\rm side}=256$ obtained through this method are available in the PLA, while for PR4 this correction has already been applied in the public polarization maps. In these public data products, the spectral index $\alpha$ has been obtained from the {\tt Commander} algorithm (see section 11.2 of \citealt{cpp2015-2}) at an effective angular resolution of $1^\circ$. Instead of using those public maps, here we choose to implement our own correction using the more precise spectral index $\alpha$ derived from our fit to the intensity SED described in section~\ref{subsec: Foreground modelling}. To this aim we downloaded from the PLA the PR3 projecting $A_Q$ and $A_U$ maps for each radiometer, and built a projection map for each frequency band as (see section 11.4 of \citealt{cpp2015-2})
\begin{equation}
P_{Q[U]} = \sum_k a_k A_{k,Q[U]}~~,
\label{eq:p_qu}
\end{equation}
where the sum extends over all radiometers in each frequency and $a_k$ is the bandpass-mismatch $a$-factor for radiometer $k$ given in Table~7 of \cite{cpp2018-2}. 

Uncertainties in this procedure have been carefully accounted for and conservatively propagated to the final error bar. We have considered the uncertainties in the estimation of the $a_k$ factors quoted in Table~7 of \cite{cpp2018-2} as well as uncertainties in the determination of the spectral index $\alpha$ that is introduced in equation~\ref{eq:qu_corr}. To this end, using equation~\ref{eq:p_qu} we have generated $P_Q$ and $P_U$ maps using the $a_k$ values corresponding to the two extremes of the error bar, i.e. $a_k-\sigma(a_k)$ and $a_k+\sigma(a_k)$ respectively, and plugged them into equation~\ref{eq:qu_corr} to produce corrected maps. Similarly, we have generated correction maps using spectral indices $\alpha-\sigma_\alpha$ and $\alpha+\sigma_\alpha$. In both cases, we calculated $Q$ and $U$ flux densities in both sets of maps, and defined two systematic uncertainties, respectively for $a_k$ and for $\alpha$, as the difference between the two extreme values. These two systematic uncertainties are added in quadrature as two additional terms in equation~\ref{eq:flux_error}. 

\begin{figure}
\begin{center}
\includegraphics[width=8.5cm]{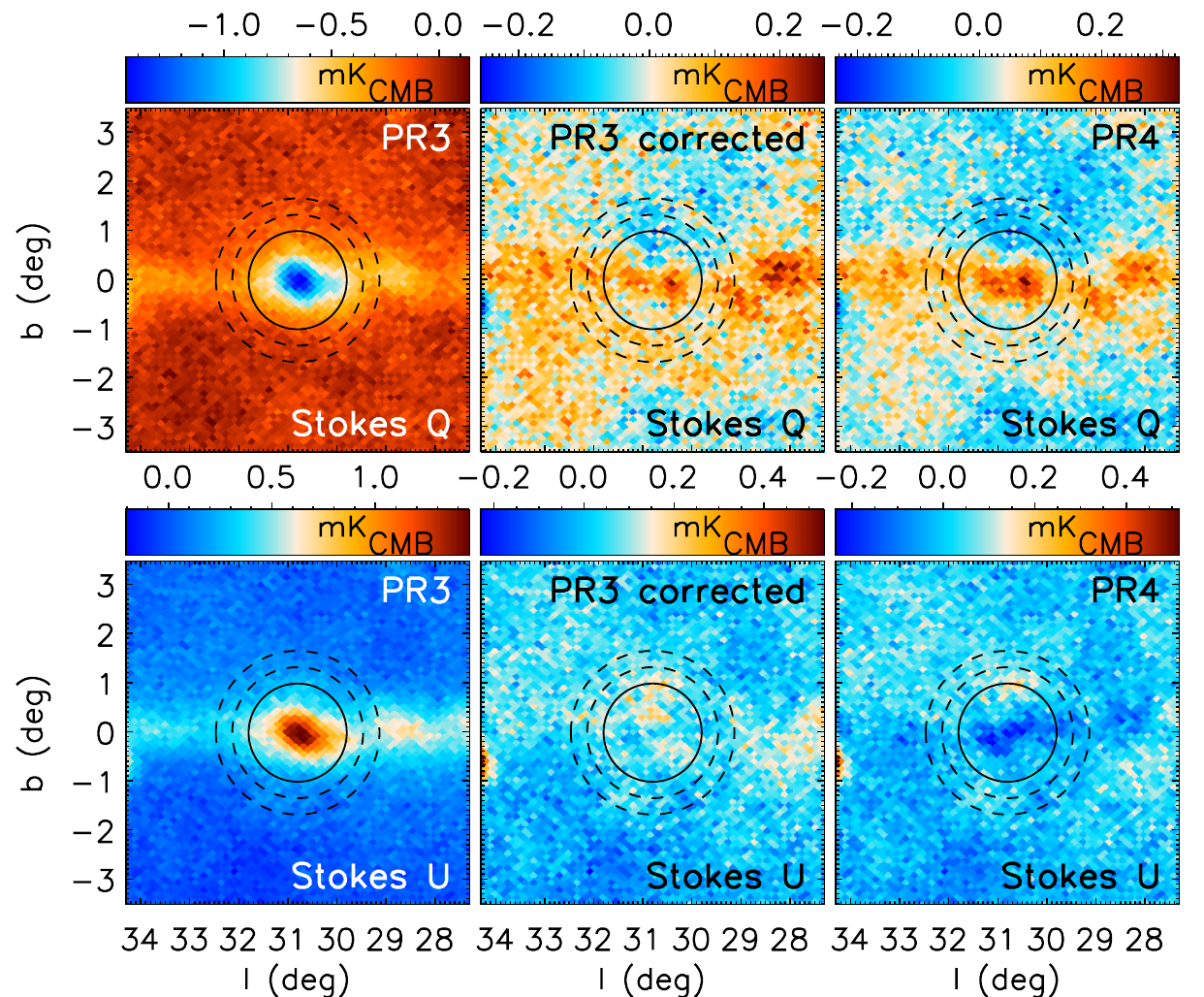}
\caption{Illustration of the effect of the polarization leakage correction in \textit{Planck}-LFI at 28.4\,GHz, around the position of W43. Stokes $Q$ and $U$ maps are represented respectively in the top and bottom panels. From left to right panels show respectively the PR3 raw (un-corrected) maps, the PR3 leakage-corrected maps using our own implementation (see Sect.~\ref{subsec:leakage_corr} for details) and the public PR4 corrected maps.}
\label{fig:W43_leakage}
\end{center}
\end{figure}

To showcase the reliability of this procedure, in Figure~\ref{fig:W43_leakage} we show the PR3 un-corrected maps, our PR3 corrected maps and the public PR4 corrected maps at 22.8\,GHz and around W43. While the un-corrected maps show significant spurious emission in $Q$ and $U$ at the position of the source, with polarization fraction of $\sim 1.5\%$, this is largely suppressed in the corrected maps. It is also clear that the PR4 maps still show some residual leakage emission, in particular in $U$, that is corrected with better accuracy in our implementation, likely thanks to a better reconstruction of the intensity spectral index that is introduced in equation~\ref{eq:qu_corr}. Diffuse emission distributed along the Galactic plane still remains in $Q$. Note also that, as in the correction procedure we have used the spectral index $\alpha$ for W43, the corrected maps are more reliable in pixels close to the central coordinates of W43 (inside the circle of Figure~\ref{fig:W43_leakage}). As we move away from the source the true underlying spectral index may deviate from that of W43, leading to a less precise correction. In any case, the leakage correction is more critical right at the position of W43 where the emission in total intensity is strong. Away from this source the emission in total intensity is much fainter so the polarization leakage is much smaller and may be embedded in the noise.

\subsubsection{Improved polarization constraints through frequency stacking}
\label{sec:stacking}
Previous similar studies have usually presented constraints on the polarization fraction of AME at individual frequencies \cite{caraballo2011,Perseus,W44}. Taking into account that the noise of data at different frequencies is statistically independent, here we consider combining the information at different frequency bands with the goal to improve the constraints on $\Pi_{\rm AME}=P/S_{\rm AME}$. This combination can be done in different ways. One possibility would be to evaluate the PDF of $\Pi_{\rm AME}$ at each individual frequency and then combine them to derive a joint constraint on $\Pi_{\rm AME}$, which is assumed here to be frequency-independent. We have implemented this method and checked that gives roughly consistent results with a different method based on a stacking at the map level that we will use as default. In this method each pixel $p$ of the stacked map is assigned a temperature value
\begin{equation}
T_p = \sum_{i} w_{i} T_{p,i} \left(\frac{\nu_{i}}{\nu_{s}}\right)^{2} \frac{\eta(\nu_i)}{\eta(\nu_s)}~~,
\end{equation}
where $T_{p,i}$ is the temperature value, in K$_{\rm CMB}$ units, of pixel $p$ at frequency $\nu_i$, $\nu_s=22.8$\,GHz is the reference frequency at which the stacking is performed, $\eta=x^2 e^x/(e^x-1)^2$ is the conversion factor between thermodynamic differential temperature (K$_{\rm CMB}$ units) and brightness Rayleigh-Jeans temperature (K$_{\rm RJ}$ units) and $w_i$ is the weight corresponding to frequency $\nu_i$. This stacking is performed independently for maps of Stokes $I$, $Q$ and $U$, using the same weights. Note that stacking both $Q$ and $U$ independently assumes that an eventual AME polarization component has a polarization angle that is constant with frequency. This assumption could be circumvented by stacking directly on polarized intensity, but at the cost of introducing additional complications related with the noise bias discussed in Sect.~\ref{sec:noise_debiasing}.

We use optimal weights to minimize the final uncertainty on $\Pi_{\rm AME}$ that then accounts not only for the uncertainties on the $I$, $Q$ and $U$ flux densities but also for the AME amplitude at each frequency $\nu_i$. In the presence of fully uncorrelated noise, these weights are given by 
\begin{equation}
w_i=\frac{I_{{\rm AME},i}^2/\sigma_i^2}{\sum_j I_{{\rm AME},j}^2/\sigma_j^2}~~.
\end{equation}
In this equation $I_{{\rm AME},i}$ represents the AME flux density at frequency $\nu_i$, calculated by subtracting from the measured flux density (calculated through equation~\ref{eq:flux_i} and listed in Table~\ref{tab:fluxes_i}) the flux densities of the sum of the rest of the components (free-free, CMB and thermal dust) resulting from our fitted model evaluated at the same frequency. The term in the denominator, $\sigma_i$, is the quadratic average of the errors of the flux-density estimates in $Q_i$ and $U_i$, $\sigma_i=\sqrt{(\sigma(Q_i)^2+\sigma(U_i)^2)/2}$.

To account for the presence of noise correlations between frequency bands, which are due to $1/f$ residuals and to background fluctuations, we use the covariance matrix in the definition of the weights, which are then given by:
\begin{equation}
w_i = \frac{\sum_j C_{ij}^{-1}\,I_{{\rm AME},i}\,I_{{\rm AME},j}}{\sum_{i,j}C_{ij}^{-1}\,I_{{\rm AME},i}\,I_{{\rm AME},j}}~~,
\end{equation}
where the sums run over frequencies, and the noise covariance matrix $C_{i,j}$ is calculated using the flux-densities calculated on the random apertures at all frequencies (see Sect.~\ref{subsec: photometry}). We calculate covariance matrices for $Q$ and $U$ independently and $C_{i,j}$ is the arithmetic mean of the two. We find strong noise correlations, of around 50-70\% for pairs of adjacent frequencies below 33\,GHz, which are driven by the background fluctuations. For instance, in W43 we find a maximum correlation of 78\% between WMAP and \textit{Planck} lowest frequency bands. \

\begin{figure}
\begin{center}
\includegraphics[width=\columnwidth]{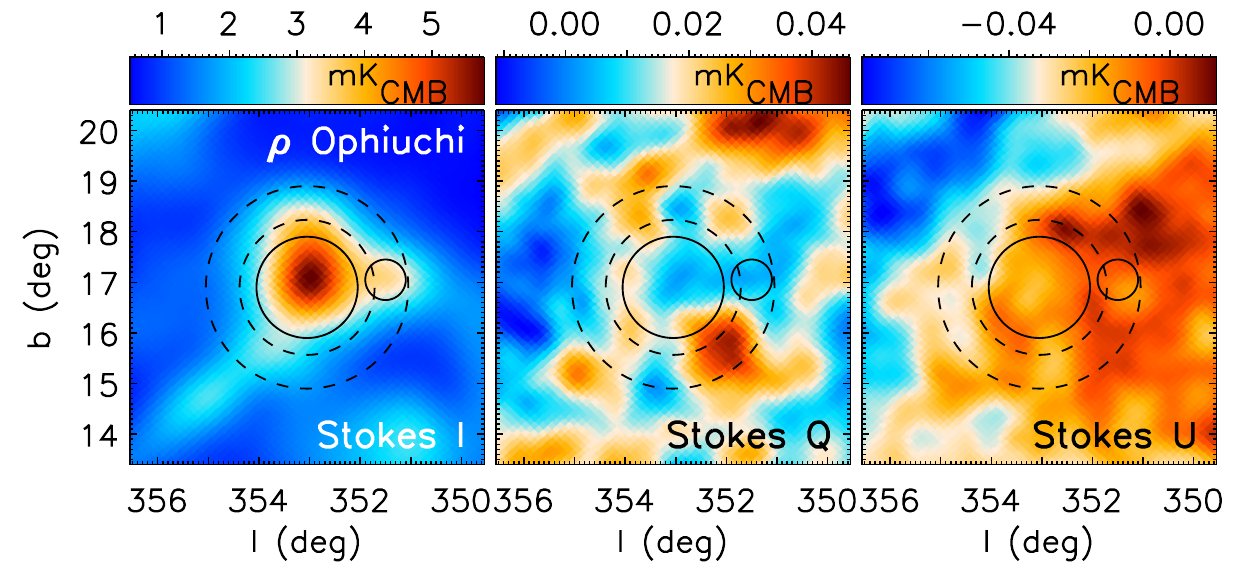}
\includegraphics[width=\columnwidth]{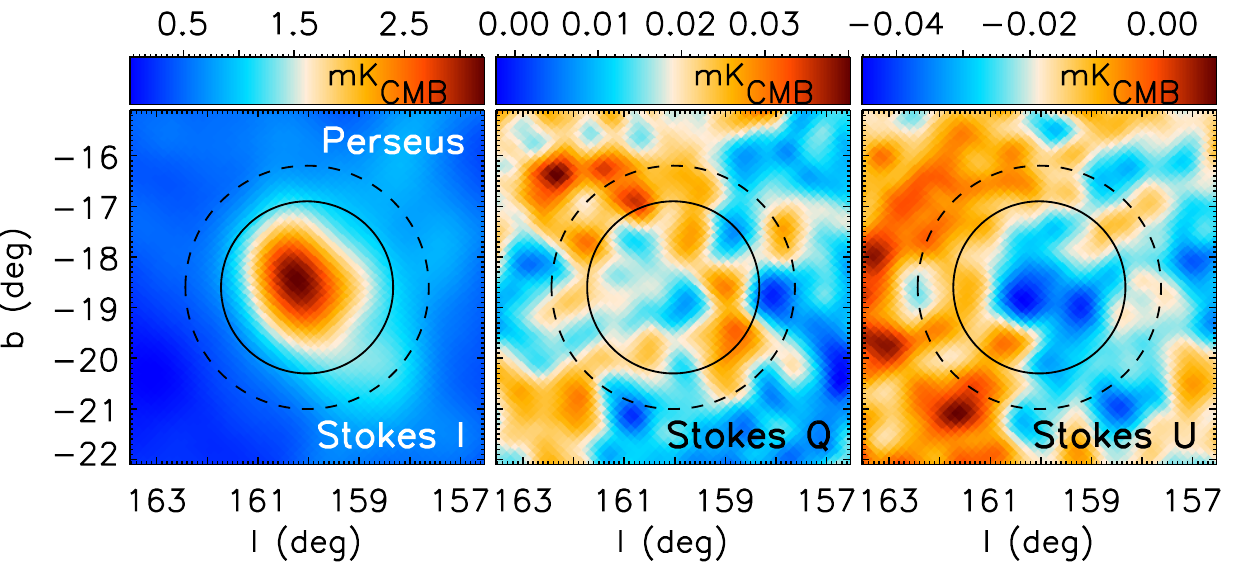}
\includegraphics[width=\columnwidth]{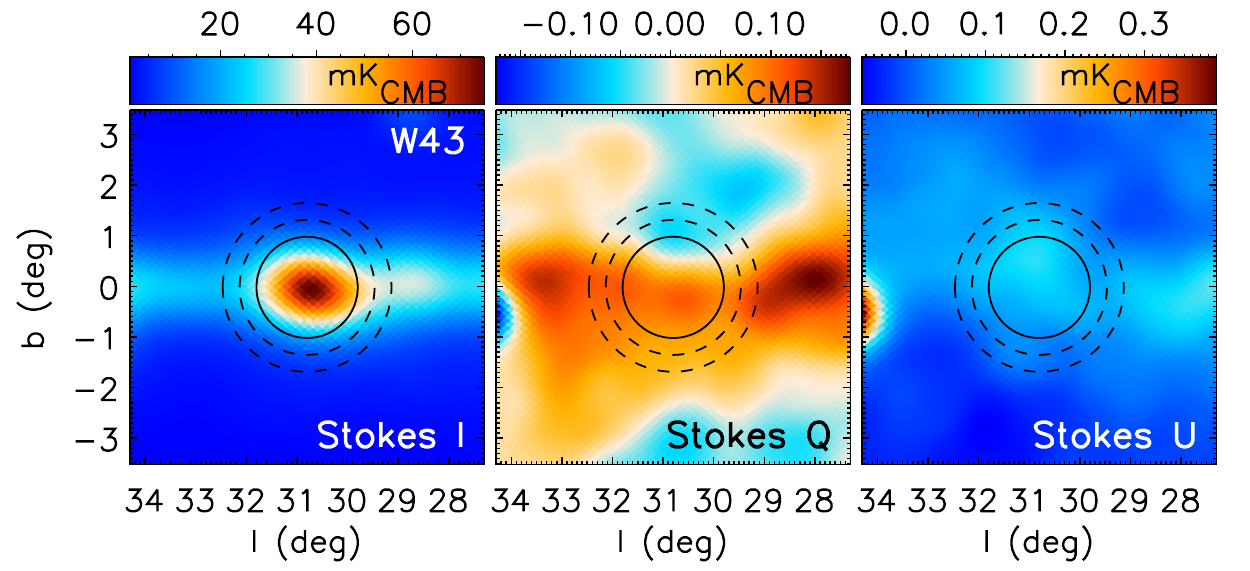}
\end{center}
\caption{Intensity and polarization stacked $I$, $Q$ and $U$ maps, at a reference frequency of 22.8\,GHz and at the position of the three sources studied in this paper. These maps are the result of a weighted average of maps at frequencies around the AME peak frequency convolved at a common angular resolution of $1^\circ$, and have been obtained following the procedure outlined in Sect.~\ref{sec:stacking}.}
\label{fig:stacked_maps}
\end{figure}

For each region we have stacked the maps corresponding to the same frequencies for which we have quoted AME polarization contraints in Tables~\ref{tab:rhoophw_pol_constraints}, \ref{tab:perseus_pol_constraints} and \ref{tab:w43_pol_constraints}. These maps have been convolved to a common angular resolution of $1^\circ$ prior to the stacking. The final stacked maps are displayed in Figure~\ref{fig:stacked_maps}. No significant emission is visible in either the $Q$ or $U$ maps except for i) diffuse emission running southwest to northeast in the $\rho$\,Ophiuchi $U$ map that is due to a large-scale synchrotron spur (see Sect.~\ref{sec:maps}), ii) diffuse emission along the Galactic plane in the $Q$ map of W43 (see Sect.~\ref{sec:maps}), and iii) polarized emission originated in the SNR W44 that is visible towards the left of the $Q$ and $U$ maps of W43. 

Flux densities are calculated on these maps through aperture photometry using equation~\ref{eq:flux_i} with the reference frequency $\nu_s=22.8$\,GHz. The residual AME flux density on the stacked map is calculated as 
\begin{equation}
S_{\rm AME,s} = S_s -\sum_i w_i(S_{\nu,i}^{\rm ff}+S_{\nu,i}^{\rm dust}+S_{\nu,i}^{\rm CMB})~~,
\end{equation}
where $S_s$ is the flux density calculated on the stacked $I$ map, and the terms inside the parenthesis are the flux densities of the different modelled components evaluated at frequency $\nu_i$. The stacked AME polarization fraction is then calculated as $\Pi_{{\rm AME},s}=\sqrt{Q_s^2+U_s^2}/I_{{\rm AME},s}$, where $Q_s$ and $U_s$ are flux densities calculated on the stacked maps, and debiased using the methodology outlined in sectcion~\ref{sec:noise_debiasing}.

\begin{table*}
\caption{Best-fitting model parameters for $\rho$\,Ophiuchi, Perseus and W43 in intensity. We compare the two cases in which we include and exclude the QUIJOTE-MFI flux densities in the fit. In the last line we show the reduced chi-squared of each fit.} 
\begin{center}
\renewcommand\arraystretch{1.3}
\begin{tabular}{lcccccc}

\hline
                &\multicolumn{2}{c}{$\rho$\,Ophiuchi} 	&\multicolumn{2}{c}{Perseus}		&\multicolumn{2}{c}{W43}  \\
Parameter 			               & With QUIJOTE	    	       & Without			         & With QUIJOTE 	      & Without                 & with QUIJOTE          & without        \\	
\hline
EM (pc$\cdot$cm$^{-6})$				    & $16^{+16}_{-11}$			   & $13^{+15}_{-9}$			&  $32 \pm 8$     	      &  $33 \pm 9$            & 3968 $\pm$ 204        & 3941 $\pm$ 200 \\ 
T$_{\rm d}$ (K)						    & $22.2 \pm ^{+0.7}_{-0.9}$	   & $22.4^{+0.6}_{-0.8}$		&  $18.6 \pm 1.1$         &  $18.7 \pm 1.1$         & 23.3 $\pm$ 1.0 	    & 23.3 $\pm$ 1.0  \\ 
$\beta_{\rm d}$						    & $1.62 \pm 0.06$			   & $1.59 \pm 0.06$ 		    &  $1.73 \pm 0.14$        &  $1.71 \pm 0.14$        & 1.69 $\pm$ 0.06	    & 1.70 $\pm$ 0.06 \\
$\tau_{250} (\times 10^{-4})$	        & $5.7^{+0.9}_{-0.5}$ 	       & $5.6^{+0.7}_{-0.4}$ 		&  $2.4^{+0.7}_{-0.5}$		  &  $2.4^{+0.7}_{-0.4} $      & 33 $\pm$ 5        & 33 $\pm$ 5 \\
$\Delta$T$_{\rm CMB} (\mu {\rm K})$		& $62 \pm 40$    			   & 52 $\pm$ 44			    &  $43 \pm 22$ 	      &  $32 \pm 24 $           & $54^{+52}_{-89}$	  & $52^{+54}_{-94}$ \\
A$_{\rm AME}$ (Jy)					    & $26.1 \pm 1.8$		       & $27.4 \pm 1.9$ 		    &  $29.4 \pm 2.6$         &  $30.5^{+3.8}_{-3.1}$   & 209 $\pm$ 19		    & 215 $\pm$ 23 \\
$\nu_{\rm AME}$ (GHz)					& $29.1^{+2.0}_{-1.5}$         & $27.4^{+2.4}_{-3.3}$		&  $25.6 \pm 1.5$ 	      &  $22.5^{+4.2}_{-4.9}$   & 24.7 $\pm$ 1.5	    & 22.9 $\pm$ 3.3 \\
W$_{\rm AME}$						    & $0.54 \pm 0.06$	           & $0.61^{+0.20}_{-0.14}$     &  $0.48 \pm 0.07$        &  $0.72 \pm 0.20$        & 0.73 $\pm$ 0.10	    & 0.82 $\pm$ 0.13 \\
\hline
$\chi^{2}_{\rm red}$                    & 0.37                         & 0.56                       & 0.12                    &  0.21                   & 1.11                  & 1.30 \\        \hline
\end{tabular}
\end{center}
\label{tab:model_parameters}
\end{table*}

\section{Results and discussion}\label{sec:results}

This section presents the main results of this paper: the modelling of the intensity SED of the three studied regions and the inferred polarization constraints for both the AME and the thermal dust emission. Figures~\ref{fig:rhoophw_sed}, \ref{fig:perseus_sed}  and \ref{fig:w43_sed} show the intensity SEDs and fitted models respectively for $\rho$\,Ophiuchi, Perseus and W43.  In Table~\ref{tab:model_parameters} we show the best-fit parameters for these three regions. To illustrate the effect of the inclusion of QUIJOTE-MFI data we also show the best-fit parameters when these data are excluded from the fit. Tables~\ref{tab:rhoophw_pol_constraints}, \ref{tab:perseus_pol_constraints} and \ref{tab:w43_pol_constraints} show the corresponding polarization constraints. In the following sections we discuss the main results for the three regions separately.

\subsection{$\rho$\,Ophiuchi} 
\label{sec:rhoophw}
\begin{figure}[!h]
\begin{center}
\includegraphics[width=1.0\linewidth,height=0.7\linewidth]{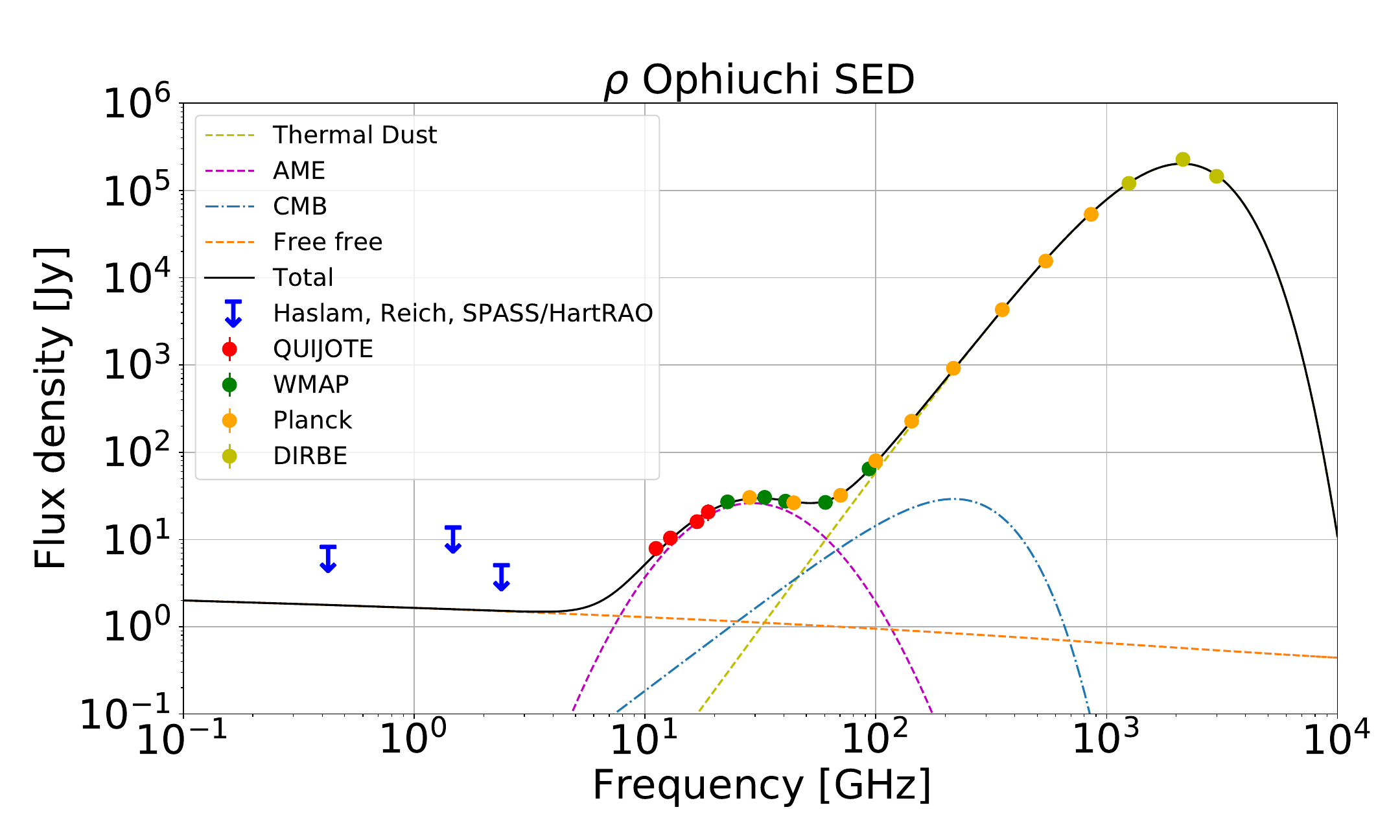}
\end{center}
\caption{$\rho$\,Ophiuchi intensity spectral energy distribution. QUIJOTE-MFI data points are depicted in red, together with other ancillary data (blue) including WMAP 9-yr (green), \textit{Planck} (orange) and COBE-DIRBE data (light green). At intermediate frequencies, the excess emission associated with the AME clearly shows up. A joint fit has been performed consisting of the following components: free-free (orange line), AME log-normal model (purple line), CMB (blue line) and thermal dust (green-olive line). The black line represents the sum of all components.}
\label{fig:rhoophw_sed}
\end{figure}

\begin{figure}[!h]
\begin{center}
\includegraphics[width=1.0\linewidth, height=0.73\linewidth]{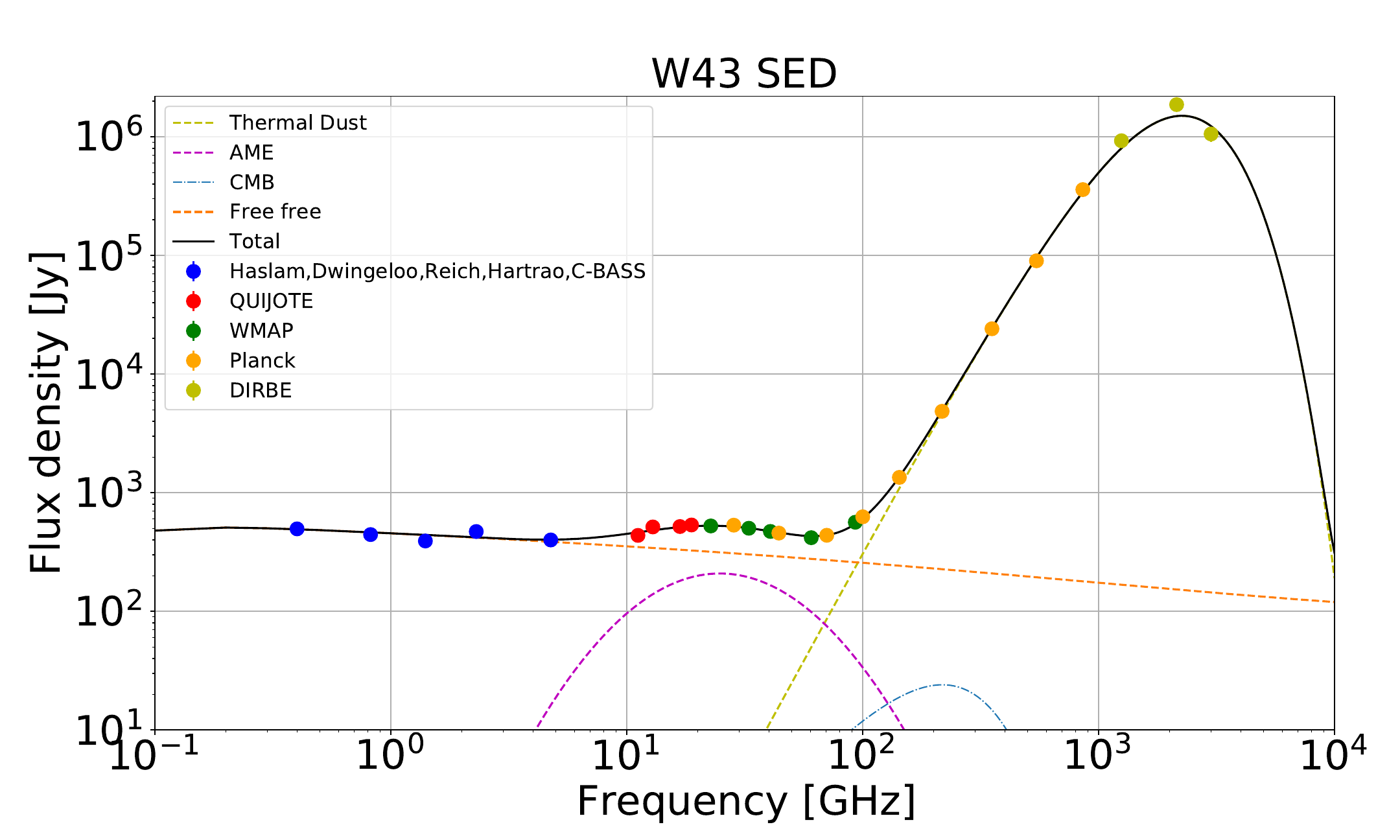}
\end{center}
\caption{Same as Figure~\ref{fig:rhoophw_sed} but for W43.}
\label{fig:w43_sed}
\end{figure}

Figure~\ref{fig:rhoophw_sed} shows the SED of the $\rho$\,Ophiuchi molecular cloud. Although the AME in this region has been extensively studied in the past \citep{casassus2008,planck2011XXnewlight}, QUIJOTE-MFI data provides, for the first time, measurements of the AME spectrum below the WMAP lowest frequency of 22.8\,GHz, as already shown in \cite{AMEwidesurvey}. Evidence for the presence of AME in this region has been solidly established for long time, as the lack of signal at low frequencies (note that all estimated flux densities below 10\,GHz are compatible with zero) is inconsistent with the flattening of the spectrum at frequencies below $\sim 60$\,GHz being due to free-free emission. Note that the three lower-frequency data points (which are depicted in Figure~\ref{fig:rhoophw_sed} as upper limits at confidence level of 95\%) were included in the fit using their central values and error bars. QUIJOTE-MFI data have allowed for the first time to delineate the downturn of the AME spectrum at low frequencies. This allows constraining the AME parameters, especially $\nu_{\rm AME}$ and $W_{\rm AME}$, with much better precision as seen in Table~\ref{tab:model_parameters}. In this case there is no improvement in the uncertainty of $A_{\rm AME}$ after the inclusion of the QUIJOTE-MFI data because the SED is markedly flat between 20 and 40\,GHz, so WMAP and \textit{Planck} data in this range are sufficient to anchor the AME amplitude. The data allow determination of the model parameters for all components with high precision, except the value of EM that is consistent with an upper limit owing to the lack of detected emission at low frequencies. These parameters are consistent with those derived in previous studies \citep{planck2011XXnewlight,AMEwidesurvey}. 

Table~\ref{tab:rhoophw_pol_constraints} shows $Q$ and $U$ flux densities, together with constraints on the polarized flux density and on the polarization fraction of the AME for frequencies below 44.1\,GHz, and for the thermal dust emission for frequencies above 60.5\,GHz. These are the first constraints on the AME polarization fraction on this region at QUIJOTE-MFI and \textit{Planck} frequencies. Note that we detect a positive signal in $U$ at frequencies up to 22.8\,GHz. As already commented by \cite{Dickinson2011} this signal is associated with a relatively bright synchrotron spur that is running diagonally across the maps. This creates a notable gradient running from southwest to northeast that is more apparent at 11\,GHz and 13\,GHz (see maps of Figure~\ref{fig:rhoophw_maps}). A fit of these $U$ values to a power-law model yields a spectral index $\alpha = -1.1 \pm 0.3$, characteristic of synchrotron emission. The signal from this spur leads to $P_{\rm db}$ values away from zero at some frequencies, degrading the upper limits on $\Pi_{\rm AME}$ shown in Table~\ref{tab:rhoophw_pol_constraints}. Yet the derived upper limit of $\Pi_{\rm AME}<1.02\%$ for \textit{Planck}-LFI 28.4\,GHz is the most stringent constraint on the AME polarization on this region; for comparison, \cite{Dickinson2011} had obtained $\Pi_{\rm AME}<1.4\%$ at 22.8\,GHz. The strongest constraint from QUIJOTE-MFI is $\Pi_{\rm AME} < 5.13\%$ at 16.8\,GHz. The combination of maps of different frequencies described in Sect.~\ref{sec:stacking} allows in this case to significantly improve the constraint, giving $\Pi_{\rm AME} < 0.58\%$. This is the most stringent upper limit on the AME polarization level ever achieved in this region.

Table~\ref{tab:rhoophw_pol_constraints} also gives values of the polarization fraction of the thermal dust emission at frequencies between 60.5\,GHz and 353\,GHz. These values are compatible with a constant value of $\Pi_{\rm dust}\approx 2\%$. We remind that these values have been obtained on maps convolved at a common angular resolution of $1^\circ$. Visual inspection of the \textit{Planck}-HFI maps at their parent angular resolution reveals inhomogeneity of the polarization direction at angular scales below $1^\circ$ and hence we conclude that the fractional polarization of the thermal dust emission is intrinsically higher at finer angular scales.

\subsection{Perseus molecular cloud}
\label{sec:perseus}

\begin{figure}[!h]
\centering
\includegraphics[width=1.0\linewidth,height=0.73\linewidth]{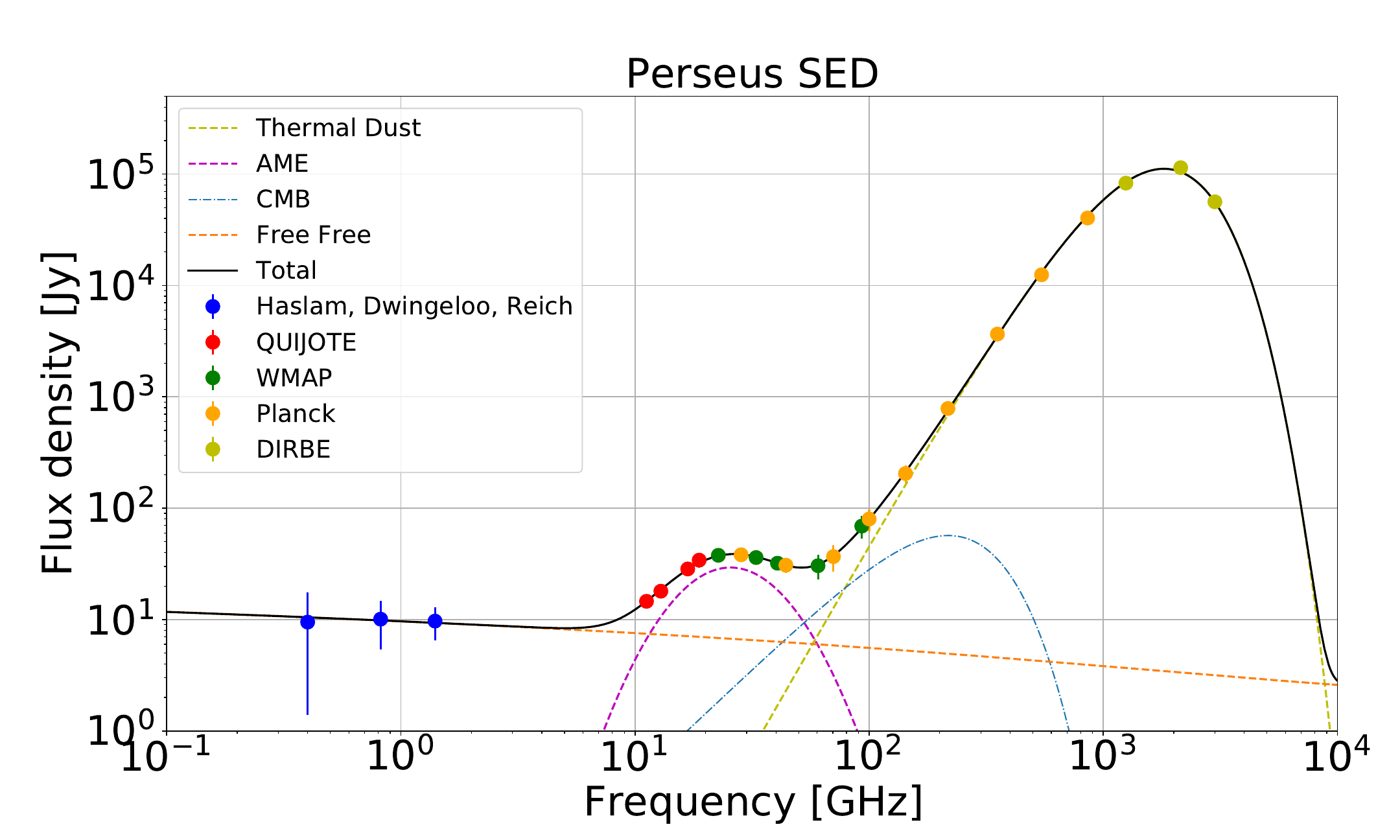}
\caption{Same as Figure~\ref{fig:rhoophw_sed} but for Perseus.}
\label{fig:perseus_sed}
\end{figure}

Figure~\ref{fig:perseus_sed} shows the SED of the Perseus molecular cloud, together with the best-fit model whose parameters are given in Table~\ref{tab:model_parameters}. It becomes clear from this table that the inclusion of the QUIJOTE-MFI data in the fit enables a more precise modelling of all AME parameters. Flux densities, as well as the best-fit model, are consistent with those derived in previous studies  \citep{Watson2005,planck2011XXnewlight,Perseus,AMEwidesurvey} in spite of small differences resulting from differences in the data analysis. QUIJOTE-MFI data in this region had already been published before \citep{Perseus}. There was also previous intensity data in the same frequency range coming from the COSMOSOMAS experiment \citep{Watson2005}. The main improvement of the data presented in this paper comes from the higher integration time per unit solid angle (see Sect.\ref{subsubsec:observation}). Yet no clear polarization signal is visible in the maps of Figure~\ref{fig:perseus_maps} nor in the stacked maps displayed in Figure~\ref{fig:stacked_maps}. 

Table~\ref{tab:perseus_pol_constraints} shows $Q$ and $U$ flux densities, together with constraints on the polarized flux density and on the polarization fraction of the AME for frequencies below 44.1\,GHz, and for the thermal dust emission for frequencies above 60.5\,GHz. As for the other two regions, errors are estimated in all cases through the scatter of the flux density values calculated on ten apertures around the source. In this particular case, the raster-scan maps have a size of $\approx 6^\circ$ (see Table~\ref{tab:obs}), and the random apertures fall in a region that, owing to not being covered by these observations, has a poorer sensitivity. To overcome this issue, we have rescaled the errors derived from the random apertures by the ratio of the pixel-to-pixel RMS calculated on the combined map (raster and nominal data) to the pixel-to-pixel RMS calculated on the map with nominal data only. Thanks to the more sensitive data, the new QUIJOTE-MFI upper limits are better by a factor $\approx 1.6$ than those presented in \cite{Perseus}. The most stringent upper limits at an individual frequency come from WMAP 22.8\,GHz and \textit{Planck}-LFI 28.4\,GHz and are similar to those obtained by \cite{caraballo2011} using WMAP 7-year data. 

The upper limit derived from the stacked maps, $\Pi_{\rm AME}<1.64\%$, is less stringent than those derived from the 22.8\,GHz and 28.4\,GHz frequency maps, even if the errors of $Q$ and $U$ are smaller in the stacked maps as expected. The reason is the $3\sigma$ detection of negative $U$ in the stacked map, which comes from the negative blue feature that is seen in the map of Figure~\ref{fig:stacked_maps}. This feature is mostly produced by the 40.6 and 44.1\,GHz maps that also give negative $U$ fluxes (see Table~\ref{tab:perseus_pol_constraints}). If we apply an alternative stacking methodology consisting in stacking the $Q$ and $U$ flux densities listed in Table~\ref{tab:perseus_pol_constraints}, using the same weights as in the map stacking, we obtain an upper limit of $\Pi_{\rm AME}<0.71\%$. This demonstrates that the final result is rather sensitive to the used methodology. Table~\ref{tab:perseus_pol_constraints} shows a polarization fraction of the thermal dust emission in the Perseus molecular cloud of $\Pi_{\rm dust}\approx 7\%$. In this case, the \textit{Planck}-HFI maps at their parent angular resolution do not show a noticeable variation of the polarization direction, so this value may be representative of the typical level of polarization in finer angular scales within this region.

\subsection{W43 molecular complex}
\label{sec:w43}

QUIJOTE-MFI maps at 16.8 and 18.8\,GHz at the position of W43 are shown in Figure~\ref{fig:w43_maps}. Owing to this region being located close to the equatorial plane, it is affected by radio-emission contamination from geostationary satellites (Figure~\ref{fig: generalmap} shows that it is very close to the masked stripe). This has left some residual contamination that is seen towards the west of these maps. This contamination is more harmful at 11 and 13\,GHz and then at these two frequencies it has only been possible to derive reliable flux densities in total intensity. 

The WMAP 22.8\,GHz map displayed in Figure~\ref{fig:w43_maps} exhibits clear diffuse emission in $Q$ along the Galactic plane. In \cite{W44} we had hypothesized that this emission could be residual free-free or AME polarization originating in W43 or diffuse synchrotron emission from the Galactic plane. While the \textit{Planck} data analysed in \cite{W44} was affected by intensity-to-polarization leakage, the improved leakage correction implemented in this paper (see Sect.~\ref{subsec:leakage_corr}) leads to a $Q$ map with a similar structure to the WMAP 22.8\,GHz. This $Q$ signal has a polarization degree of $\approx 0.3\,\%$ at 22.8\,GHz. The similarity of the WMAP and \textit{Planck}-LFI maps at these two frequencies, and the behaviour of the flux densities in $Q$ at higher frequencies showing a monotonic decrease (see Table~\ref{tab:w43_pol_constraints}) could naturally lead to the conclusion that this is a real signal and that the leakage is controlled to levels of $\sim 0.2\%$ or better. On the contrary, the signal in $U$ shows a different behaviour, with variations in sign at a level larger than the uncertainty and with amplitude of $\sim 0.2\%$, pointing to the presence of possible leakage residuals or any other unaccounted systematic effects at this level. This also becomes evident in the comparison of the WMAP and \textit{Planck}-LFI $U$ maps shown in Figures~\ref{fig:w43_maps} and \ref{fig:W43_leakage}, which show different structure. We have performed a joint fit of the $Q$ and $U$ values to a power-law (common spectral index and different amplitudes in $Q$ and $U$) that gives $\alpha=-1.47\pm 0.94$, a spectral index that is consistent with synchrotron emission. However, this fit has $\chi^2=35.4$ with 15 degrees of freedom, tentatively pointing to an underestimation of the uncertainties. We have undertaken a detailed study on bright unpolarized sources that shows that the residual intensity-to-polarization leakage in \textit{Planck}-LFI is at a level below $0.2\%$ (see appendix \ref{appendix}). This is precisely of the same order of the signals in $Q$ and $U$ in W43. Therefore, we believe that with the current data it is not possible to claim that the signal in $Q$ is real, even if the frequency spectrum traced by three different experiments could be suggestive that there could be some contribution from diffuse synchrotron emission or even possibly from the AME originating in W43. Disentangling between these hypotheses would require data in the same frequency range but with a control of systematic effects below the $0.2\%$ level. This is a goal for the QUIJOTE TFGI instrument operating at 30 and 40\,GHz. Future polarization data from C-BASS at 5\,GHz in this region will also be very useful, in particular to test the synchrotron hypothesis.

Given the ambiguity on the interpretation of the origin of the $Q$ signal in W43, we have decided to quote upper limits on the polarization fraction of AME, as shown in Table~\ref{tab:w43_pol_constraints}. We have obtained $\Pi_{\rm AME}<0.28\,\%$ at 33.0\,GHz. This region gives the most stringent constraints on the level of AME polarization ever achieved. In \cite{W44} we had obtained $\Pi_{\rm AME}<0.22$\,\% at 40.6\,GHz. The reason why the constraint quoted in Table~\ref{tab:w43_pol_constraints} is looser is differences in the intensity modelling of AME that lead to a lower residual AME flux density at this frequency. The stacked maps displayed in Figure~\ref{fig:stacked_maps} also show a positive signal in $Q$, and lead to a constraint of $\Pi_{\rm AME}<0.31$\,\%. Similarly to what happened in Perseus (see previous section), the stacking procedure does not lead to a more stringent upper limit because it is affected by the positive $Q$ signal that is measured at individual frequencies. In other words, the stacking reduces the uncertainty on the measurement of this positive signal, but the upper limit is not affected because it depends on the central value and not on its uncertainty. This leads us to the conclusion that any future improvement on the derived upper limits depends more on a better understanding of the residual polarization signals that are seen in Perseus and in W43 (potentially through data in a different frequency range) than on improving the sensitivity. Table~\ref{tab:w43_pol_constraints} also lists the polarization degree of thermal dust emission at frequencies above 93.5\,GHz, which has a value of $\Pi_{\rm dust}\approx 1\%$. In this case, inspection of the \textit{Planck}-HFI maps at their parent angular resolution reveals a notable spatial variability of the polarization direction inside the $1^\circ$ circular aperture, and then the intrinsic polarization fraction in compact regions inside this aperture may be higher.

\section{Conclusions} 
\label{sec:conclusions}
We have presented a joint study of the microwave AME emission, with emphasis on their polarization properties, of three of the brightest or best characterized AME regions on the sky: the $\rho$ Ophiuichi and Perseus molecular clouds and the W43 molecular complex. This study focuses on the use of new or improved data from the QUIJOTE-MFI instrument at 11, 13, 17 and 19\,GHz, which crucially help to better trace the low-frequency tail of the AME spectrum and to add further constraints on the AME polarization at these frequencies. With respect to previous QUIJOTE studies on Perseus \citep{Perseus} and on W43 \citep{W44}, we have included new data and have implemented an improved calibration and data processing, which has allowed reaching in these regions sensitivity levels in polarization in the range $10-50~\mu$K\,deg$^{-1}$ depending on the frequency. The Perseus field is amongst the ones with higher integration time per unit area of all fields observed with QUIJOTE-MFI. The QUIJOTE-MFI data have provided, for the first time, the detection of emission from the $\rho$\,Ophiuchi molecular cloud below 20\,GHz, and hence has allowed the first unambiguous characterization of the AME spectrum below its peak in this region. In this paper we have also presented the first constraints on the level of AME polarization after applying an improved intensity-to-polarization leakage correction of \textit{Planck}-LFI data. This is based on the implementation of a careful correction of the intensity-to-polarization leakage of \textit{Planck}-LFI data, one of the most harmful systematic effects of these data that, if uncorrected, renders these data useless for any reliable analysis especially in bright regions such as W43. This correction critically depends on an accurate characterization of the intensity spectrum of the source, and we demonstrate that we have achieved a more reliable local correction using \textit{Planck} PR3 data than what has been implemented in the PR4 maps. One further novelty of this paper is the application of a combination of all frequency bands sensitive to AME with the goal to improve the final constraint on the AME polarization under the assumption that data at different frequencies are statistically independent.

We have fitted the AME intensity spectra using a 3-parameter phenomenological model consisting in a parabola in the log-log plane, which for certain combination of parameters allows us to reproduce accurately typical spinning dust spectra. As anticipated, the inclusion of the QUIJOTE-MFI data at frequencies 10--20\,GHz helps to better constrain the parameters of this model, in particular its peak frequency ($\nu_{\rm AME}$) and width ($W_{\rm AME}$), whose errors decrease by a factor $2$--$3$. This improved characterization of the AME intensity spectrum is critical to derive more reliable constraints on the AME polarization level. In $\rho$\,Ophiuchi we determine $\Pi_{\rm AME}<1.02\%$ (95\% C.L.) from \textit{Planck}-LFI at 28.4\,GHz that is the most stringent constraint on the AME polarization level in this region, slightly improving previous results in the same region \citep{Dickinson2011}. The most stringent constraint from QUIJOTE-MFI in this case is $\Pi_{\rm AME} < 5.13\%$ at 16.8\,GHz, while stacking all frequencies between 11.1 and 44.1\,GHz leads to $\Pi_{\rm AME}<0.58\%$. This is the second best constraint ever achieved on an individual region after W43. The new QUIJOTE-MFI data on Perseus allowed achieving $\Pi_{\rm AME}<3.5\%$ at 16.8 and 18.8\,GHz, which represents an improvement of $\approx 35\%$ with respect to the results presented in \cite{Perseus}. At other frequencies the constraints in Perseus are similar or slightly better than those derived in previous publications \citep{caraballo2011,Dickinson2011}, with a best constraint of $\Pi_{\rm AME}<1.09\%$ from WMAP 22.8\,GHz. The constraints derived from W43 are slightly looser than those presented in \cite{W44} because differences in the calibration and data processing have led to lower AME residual flux densities in total intensity. Here we obtain a best upper limit of $\Pi_{\rm AME}<0.29\%$ from WMAP 33.0\,GHz. The frequency-stacking technique in Perseus and in W43 leads respectively to $\Pi_{\rm AME}<1.64\%$ and $\Pi_{\rm AME}<0.31\%$. The reason why the stacking in these two cases does not lead to better constraints than the best constraint achieved at an individual frequency is that in both cases they are driven not by statistical errors in the data but by a detection of residual polarization emission. Improving these constraints requires a better understanding of the nature of this emission. This could be achieved through more sensitive data at various frequencies, and ideally with finer angular resolution, in order to enable a precise characterization of their spectra. In the case of W43, given that the measured signal in $Q$ is at level of $0.2\%$ with respect to the measured intensity, it would also be needed a control of instrument systematics at this level, something that is hard to achieve. We have pushed the current data to their limits and improving the AME polarization constraints in W43 would then require further technical and observational efforts.

The constraints on AME polarization presented in this paper are amongst the most stringent achieved on compact regions. They benefit from very low or no free-free emission on the $\rho$\,Ophiuchi and Perseus molecular clouds which are both located away from the Galactic plane. On the other hand W43 has significant free-free emission whose level is nevertheless relatively well anchored by the low-frequency data. These results are important not only to shed information on different AME models but to assess up to what extent AME could be a problem for the search of the B-mode signal in the CMB polarization. It must be borne in mind however that here we have analyzed three regions with specific physical conditions, so these results cannot be easily generalized. Additional constraints on AME polarization in large portions of the sky are therefore needed, especially in what concerns primordial B-mode studies. On the other hand, it must be noted also that these constraints have been obtained at an angular scale of $1^\circ$. In what concerns B-mode searches this might be sufficient, as this signal shows up at large angular scales. However, 
in order to provide useful feedback on AME modelling, it would be important to derive constraints on the AME polarization at finer angular scales so to avoid beam-depolarization effects.

\section*{Acknowledgements}
\input{quijote_acknow}

This research made use of computing time available on the high-performance computing systems at the IAC. We thankfully acknowledge the technical expertise and assistance provided by the Spanish Supercomputing Network (Red Espa\~nola de Supercomputaci\'on), as well as the computer resources used: the Deimos/Diva Supercomputer, located at the IAC. This research used resources of the National Energy Research Scientific Computing Center, which is supported by the Office of Science of the U.S. Department of Energy under Contract No. DE-AC02-05CH11231. RGG acknowledges support from Italian Ministry of education, university and research. MFT acknowledges support from from the Agencia Estatal de Investigaci\'on (AEI) of the Ministerio de Ciencia, Innovaci\'on y Universidades (MCIU), from the European Social Fund (ESF) under grant with reference PRE-C-2018-0067 and from the French Programme \textit{d'investissements d'avenir} through the Enigmass Labex. FP acknowledges support from the Spanish Ministerio de Ciencia, Innovación y Universidades (MICINN) under grant numbers PID2022-141915NB-C21. DT acknowledges financial support from the XJTLU Research Development Fund (RDF) grant with number RDF-22-02-068. CHM acknowledges financial support from the Spanish Ministry of Science and Innovation under project PID2021-126616NB-I00.  This research has made use of the SIMBAD database, operated at CDS, Strasbourg, France \citep{simbad}. Some of the presented results are based on observations obtained with \textit{Planck} ({\tt http://www.esa.int/Planck}), an ESA science mission with instruments and contributions directly funded by ESA Member States, NASA, and Canada. We acknowledge the use of the Legacy Archive for Microwave Background Data Analysis (LAMBDA). Support for LAMBDA is provided by the NASA Office of Space Science.  Some of the results in this paper have been derived using the healpy and {\tt HEALPix} packages \citep{Healpix, Healpix2}. We have also used {\tt scipy} \citep{scipy}, {\tt emcee} \citep{emcee}, {\tt numpy} \citep{numpy}, {\tt matplotlib} \citep{matplotlib}, {\tt corner} \citep{corner} and {\tt astropy} \citep{astropy1,astropy2} \textsc{python} packages.

\bibliographystyle{mnras}
\bibliography{quijote,quijote_ame_pol_3r}

\appendix 
\section{Assessment of the level of intensity-to-polarization leakage in WMAP and \textit{Planck}-LFI}
\label{appendix}

Pushing down the upper limits on the AME polarization degree requires a careful characterization of other physical mechanisms producing polarized emission, either Galactic emission (synchrotron emission in particular) or instrumental effects. In the brightest source in our sample, W43, we have detected a signal with a polarization degree of $\approx 0.2\%$ whose origin is not clear (see discussion in section~\ref{sec:w43}). In part motivated by the need to understand the origin of this signal, in this appendix we present a detailed study of the level of polarization leakage in QUIJOTE-MFI, WMAP and \textit{Planck}-LFI. To this aim we analyse the polarization data in three bright un-polarized regions: the SNR Cas\,A and the HII regions M42 (also known as ``Orion nebula'') and Cygnus\,X. Despite being a SNR dominated by synchrotron emission in the radio domain, Cas\,A is known to be largely depolarized due to the combination of various effects the most important of which is internal Faraday depolarization \citep{Anderson_95}. On the other hand, the radio emission of M42 and Cygnus\,X is fully dominated by free-free that is intrinsically unpolarized.

\begin{figure*}
\includegraphics[width=18.8cm]{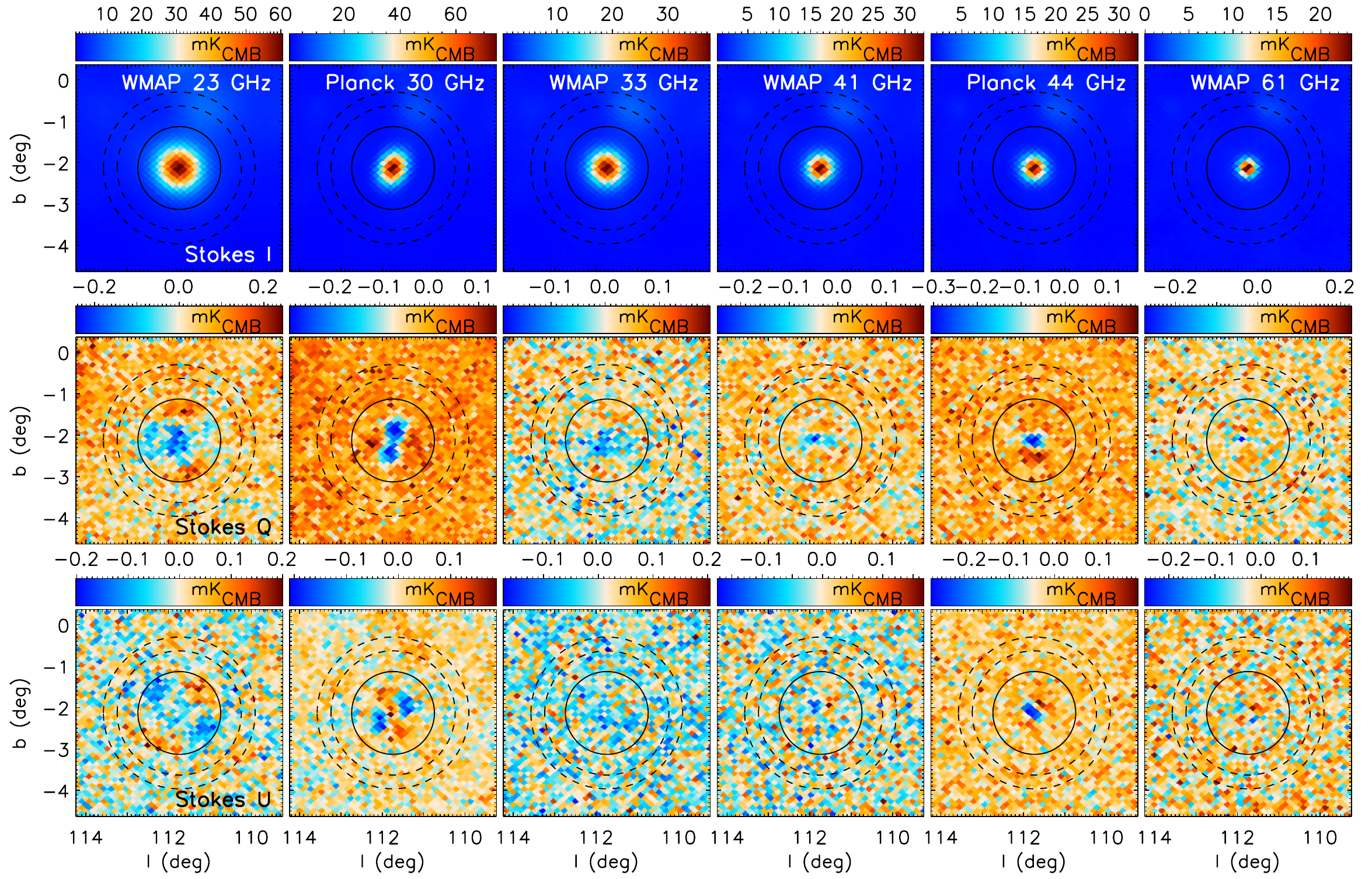}
\caption{WMAP and \textit{Planck}-LFI total intensity (top row) and polarization (middle row Stokes $Q$, bottom row Stokes $U$) maps at the position of the SNR Cas\,A. The $Q$ and $U$ polarization maps show the cloverleaf-shaped pattern typical of beam polarization. The solid and dashed circles denote the regions we have used for aperture photometry integration and background subtraction to derive the values quoted in Tables~\ref{tab:wmap_leakage} and \ref{tab:lfi_leakage}.}
\label{fig:leakage_cass}
\end{figure*}

\begin{figure*}
\includegraphics[width=18.8cm]{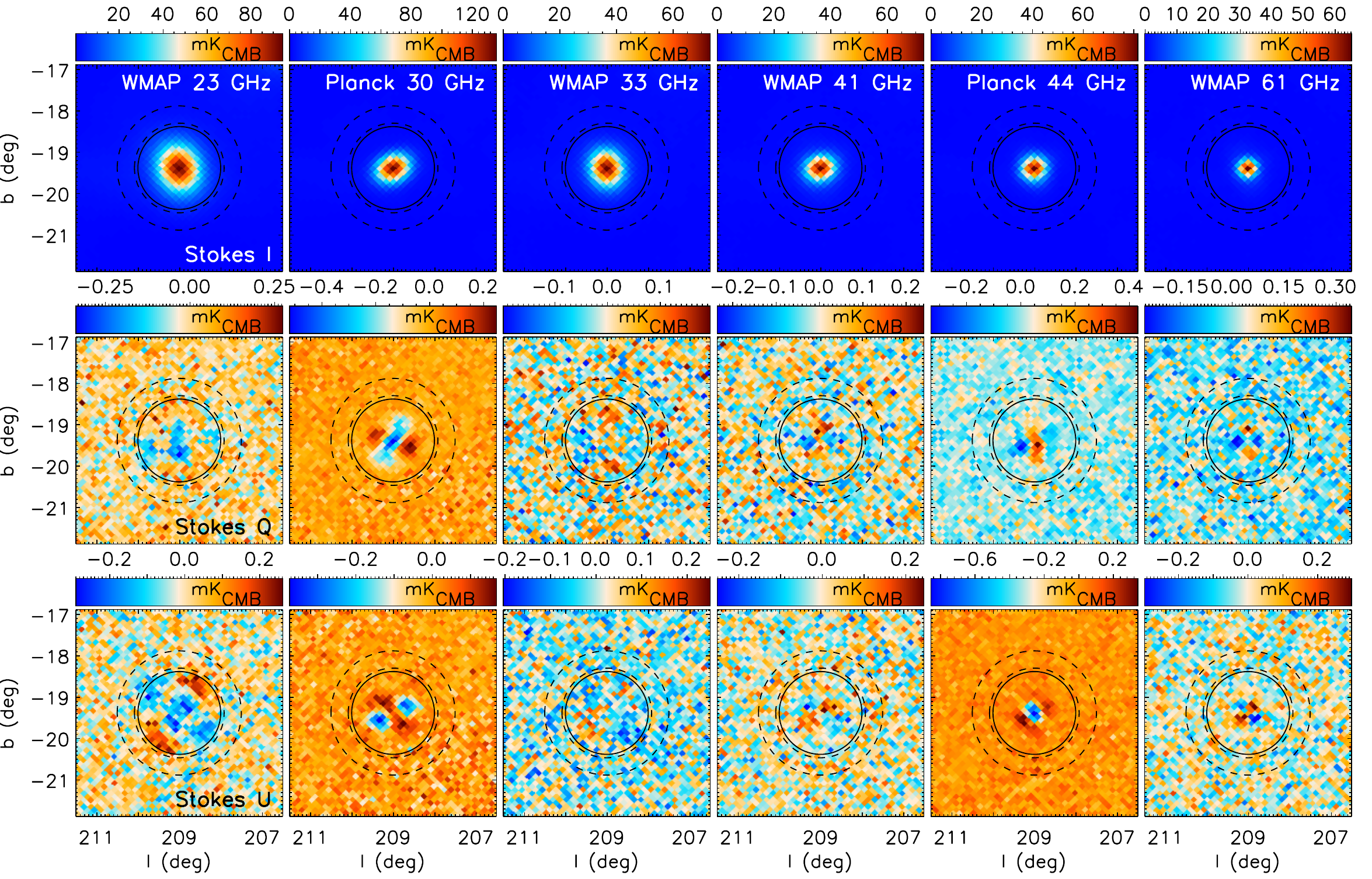}
\caption{WMAP and \textit{Planck}-LFI total intensity (top row) and polarization (middle row Stokes $Q$, bottom row Stokes $U$) maps at the position of the HII region M42. As in Figure~\ref{fig:leakage_cass}, the $Q$ and $U$ polarization maps show the cloverleaf-shaped pattern typical of beam polarization. The solid and dashed circles denote the regions we have used for aperture photometry integration and background subtraction to derive the values quoted in Tables~\ref{tab:wmap_leakage} and \ref{tab:lfi_leakage}.}
\label{fig:leakage_m42}
\end{figure*}

\begin{figure*}
\includegraphics[width=18.8cm]{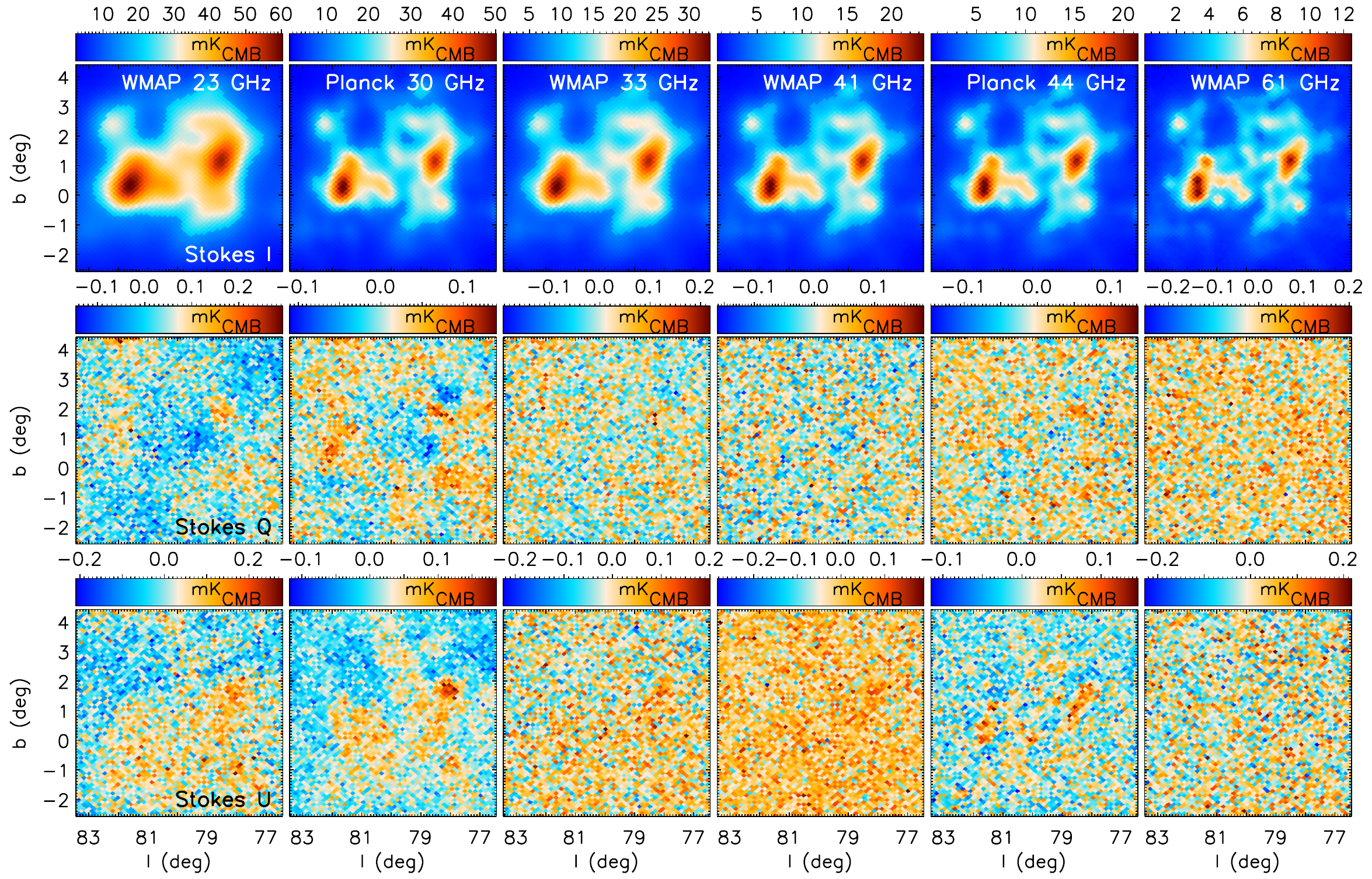}
\caption{WMAP and \textit{Planck}-LFI total intensity (top row) and polarization (middle row Stokes $Q$, bottom row Stokes $U$) maps at the position of the Cygnus\,X molecular complex. Some Stokes $Q$ and $U$ maps show residual polarization probably associated with spurious intensity leakage produced by bandpass missmatch. The unshaded regions in the top row indicate the pixels that are used in the correlation plot analysis (see the main text for details).}
\label{fig:leakage_cygnus}
\end{figure*}

For \textit{Planck}-LFI we have applied the correction procedure explained in section~\ref{subsec:leakage_corr}. To that aim for Cas\,A we have used a spectral index $\alpha=-0.71$ \citep{Weiland_11}. In the case of M42 and Cygnus\,X we have used $\alpha=-0.131$, -0.138 and -0.144 respectively for the 28.4, 44.1 and 70.4\,GHz frequency bands. These indices are derived from a fit to a power-law spectrum of the free-free spectrum given in equation~\ref{eq:sff}, using the optical depth and Gaunt factor given by equations~\ref{eq:tauff} and \ref{eq:gaunt_factor} respectively.

Figures~\ref{fig:leakage_cass}, \ref{fig:leakage_m42} and \ref{fig:leakage_cygnus} present WMAP and \textit{Planck}-LFI maps of Stokes $I$, $Q$ and $U$ of these three regions. Being compact sources, with an angular extent much smaller than the beam width, the polarization maps of Cas\,A and M42 show the typical cloverleaf pattern with two positive lobes and two orthogonal negative lobes. This is produced by the well-known ``beam mismatch'', which is the difference in the copolar beams of the two radiometers that measure the two orthogonal polarizations (neglecting the cross-polar terms contributions, that are known to be small). Peak-to-peak this signal is found to be of order $\approx 0.2-0.4\%$ in WMAP and \textit{Planck}-LFI, while in QUIJOTE-MFI is $\lesssim 1\%$ \citep{mfiwidesurvey}. We have performed a quantitative analysis by applying an aperture photometry integration around the position of these sources. From the derived intensity and polarization flux densities we have derived the $Q/I$ and $U/I$ polarization fractions listed in Tables \ref{tab:wmap_leakage} and \ref{tab:lfi_leakage}. For comparison in Table~\ref{tab:lfi_leakage} we also show the polarization degrees derived from the PR3 un-corrected maps and from the PR4 maps. Obviously, in an aperture integration the positive and negative structures will partially cancel out giving a smaller polarization percentage. We find typically $Q/I$ and $U/I$ values below 0.5\%. Given that the analyses presented in this paper are based on aperture-photometry integrations, these values give a reference of the level by which our analyses may be affected by beam mismatch. Note however that the intensity emission of the three regions analysed in this paper is mostly extended so beam effects might be largely reduced.

\begin{figure}
\includegraphics[width=\columnwidth]{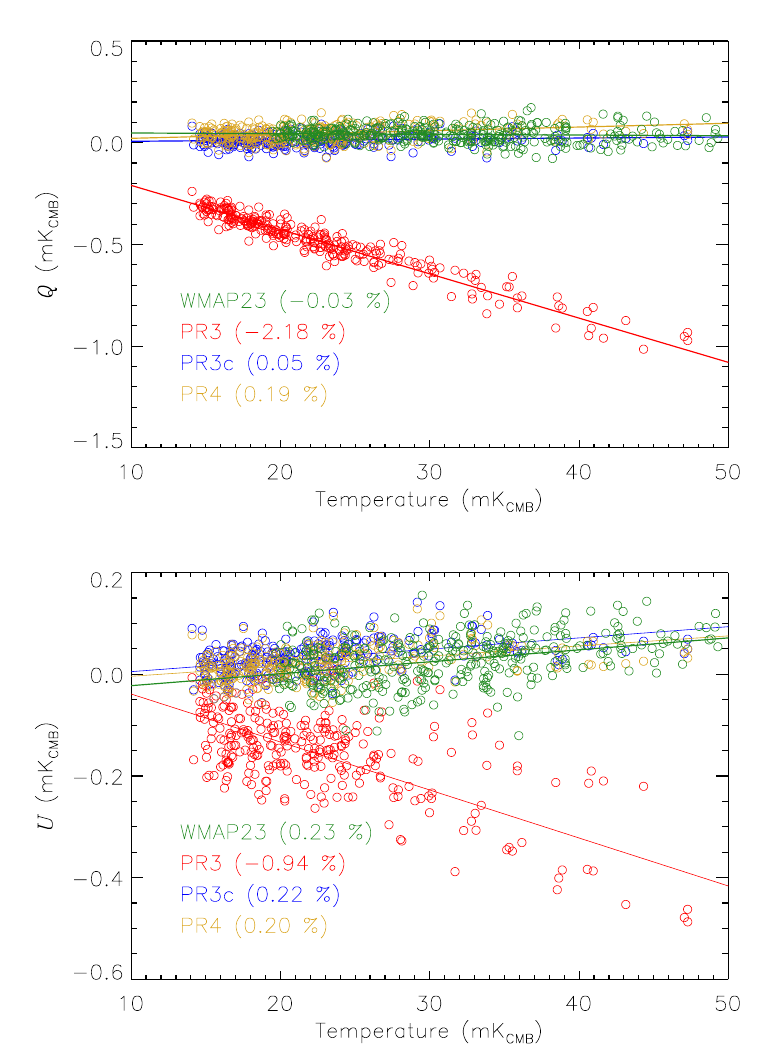}
\caption{Stokes parameters $Q$ (top) and $U$ (bottom) versus total intensity signal in the Cygnus\,X star-forming complex. With different colours we represent WMAP 23\,GHz (green) and \textit{Planck}-LFI 30\,GHz data for three different cases: PR3 raw (un-corrected) data (red), PR3 leakage-corrected data (blue) and PR4 leakage-corrected data (gold). In the legend we quote the $Q/I$ and $U/I$ polarization fractions derived from the linear-regression fits represented by the solid lines (same values that are quoted in Tables~\ref{tab:lfi_leakage} and \ref{tab:wmap_leakage}.).}
\label{fig:qu_vs_i_cygnus}
\end{figure}

On the contrary the intensity emission in Cygnus\,X extends mostly on angular scales larger than the beam, and so in this case beam mismatch is expected to be reduced. The WMAP and \textit{Planck}-LFI polarization maps exhibit however emission with some spatial resemblance. In particular, the positive feature around $(l,b)\approx (78^\circ,2^\circ)$ seems to be present in all the five lower-frequencies $U$ maps. A positive signal roughly at the same position is seen in the 23 and 30\,GHz $U$ maps, with a negative structure to the south. While both WMAP and \textit{Planck}-LFI can potentially suffer from leakage associated with bandpass mismatch\footnote{The subtraction of the signals measured by the two radiometers measuring the two orthogonal polarization contains some residual intensity signal when the two bandpasses have different spectral shapes.} that could lead to spatially-correlated polarized structure, coincidence in polarization direction across frequency bands is hard to be explained by this effect. On the contrary, given that the emission in Cygnus\,X is found to be free-free dominated, it is hard to think of any real polarization signal. We prefer to adopt an aseptic position and, assuming that this signal is produced by intensity-to-polarization leakage, will infer an upper limit on the polarization degree of this effect. In order to estimate the polarization degree of this signal, we have performed a correlation-plot analysis in which we represent $Q$ and $U$ versus the total intensity of each pixel. To reduce correlations between pixels we first degrade the maps to $N_{\rm side}=256$. We then perform a fit to a linear polynomial of all pixels with total-intensity values above a given threshold (the resulting masks can be seen in the top row of Figure~\ref{fig:leakage_cygnus}), whose slopes give an estimate of the average polarization degrees in terms of $Q/I$ and $U/I$. Figure~\ref{fig:qu_vs_i_cygnus} shows an example of these fits for the 23\,GHz and 28.4\,GHz bands of WMAP and \textit{Planck}-LFI, which show that the un-corrected PR3 data has a leakage level of up to $\approx 2\%$. The fitted slopes for all bands of WMAP and \textit{Planck}-LFI are given in Tables~\ref{tab:wmap_leakage} and \ref{tab:lfi_leakage}. We have applied the same methodology to QUIJOTE-MFI and obtained consistent results with those reported in \cite{mfiwidesurvey}, concluding that the leakage is at level $\lesssim 1\%$. In WMAP and \textit{Planck}-LFI the fitted slopes in $Q/I$ are mostly consistent with zero (except at 70.4\,GHz) in spite of the presence of positive and negative structures in the $Q$ maps (see Figure~\ref{fig:leakage_cygnus}) that could actually partially cancel out. In this sense, given the proximity of these regions on the sky the projection maps of equation~\ref{eq:qu_corr} will differ very little, and then a change in the sign the leaked $Q$ (or $U$) signal can only be explained through a significant change of the spectral index $\alpha$. The fact that the intensity emission in Cygnus\,X is largely dominated by free-free, which has a well-defined spectral index, renders this hypothesis rather unplausible. The only possible remaining hypotheses are then either the presence of beam mismatch leakage or that this signal is real. The $U$ maps on the contrary show mostly positive structure giving a slope of $\approx 0.2\%$ with remarkable consistency in the four lower-frequency maps (WMAP 22.8, 33.0 and 40.6\,GHz and \textit{Planck}-LFI 28.4\,GHz). Spatial resemblance of the $Q$ emission on different frequencies, and the consistency of the $Q/I$ level strengthens the idea that this signal is real.

As a summary of the analyses presented in this appendix, given the possible presence of real polarization signal, even if at a very low level, rather than fixing the polarization leakage at a certain level it seems more reliable to quote an upper limit. In this sense it seems robust to conclude that the intensity-to-polarization leakage is below the 1\% level in QUIJOTE-MFI and below the 0.2\% level in \textit{Planck}-LFI and WMAP.

\begin{table*}
\caption{Level of leakage in WMAP, computed on three bright unpolarized regions. We quote polarization fractions $Q/I$ and $U/I$ for each of the five WMAP frequencies.}
\begin{center}
\begin{tabular}{ccccccccc}
\hline
  Freq.  &  \multicolumn{2}{c}{Cas\,A}       &&     \multicolumn{2}{c}{M42}        &&     \multicolumn{2}{c}{Cygnus\,X}   \\
\cline{2-3}\cline{5-6}\cline{8-9}
  (GHz)  &  Q/I (\%)         &      U/I (\%)       &&     Q/I (\%)        &     U/I (\%)         &&     Q/I (\%)       &     U/I (\%)   \\
\cline{2-9}
  22.8  &  $-0.28 \pm 0.04$  &   $-0.13 \pm  0.04$ &&  $-0.12 \pm  0.03$  &   $-0.07 \pm  0.03$  &&  $-0.03 \pm 0.02$  &  $ 0.23 \pm 0.03$   \\
  33.0  &  $-0.25 \pm 0.09$  &   $-0.11 \pm  0.10$ &&  $-0.04 \pm  0.06$  &   $ 0.02 \pm  0.05$  &&  $-0.08 \pm 0.03$  &  $ 0.23 \pm 0.04$   \\
  40.6  &  $-0.39 \pm 0.15$  &   $-0.08 \pm  0.14$ &&  $ 0.05 \pm  0.10$  &   $-0.06 \pm  0.09$  &&  $ 0.08 \pm 0.05$  &  $ 0.23 \pm 0.05$   \\
  60.5  &  $ 0.08 \pm 0.53$  &   $-0.18 \pm  0.48$ &&  $-0.07 \pm  0.21$  &   $ 0.16 \pm  0.21$  &&  $ 0.10 \pm 0.11$  &  $ 0.04 \pm 0.12$   \\
  93.5  &  $ 2.25 \pm 1.48$  &   $-2.79 \pm  1.34$ &&  $-0.32 \pm  0.43$  &   $ 0.22 \pm  0.49$  &&  $-0.46 \pm 0.22$  &  $-0.05 \pm 0.25$   \\
\hline
\end{tabular}
\end{center}
\label{tab:wmap_leakage}
\end{table*}

\begin{table*}
\caption{Level of leakage in \textit{Planck}-LFI maps, computed on three bright unpolarized regions. We quote polarization fractions $Q/I$ and $U/I$ for each of the three \textit{Planck}-LFI frequencies. We show results obtained on the PR3 maps, on the PR3 leakage-corrected maps and in the PR4 maps.}
\begin{center}
\begin{tabular}{cccccccc}
\hline
 Freq. & \multicolumn{3}{c}{$Q/I$ (\%)} && \multicolumn{3}{c}{$U/I$ (\%)}\vspace{0.1cm} \\
\cline{2-4}\cline{6-8}
 \vspace{0.1cm}
(GHz) &           PR3       &        PR3c       &       PR4     &~&            PR3      &          PR3c    &  PR4    \\ 
\hline
 & \multicolumn{7}{c}{Cas\,A}\\
28.4 &      $ 1.38\pm 0.07$ &  $-0.34\pm 0.05$  & $-0.03\pm 0.04$   &&    $-2.39\pm 0.07$ &  $-0.13\pm 0.04$ &  $-0.80\pm 0.04$   \\
44.1 &      $-0.02\pm 0.17$ &  $-0.26\pm 0.17$  & $-0.06\pm 0.14$   &&    $ 0.02\pm 0.15$ &  $-0.06\pm 0.15$ &  $-0.16\pm 0.13$   \\
70.4 &      $-0.53\pm 0.43$ &  $-0.13\pm 0.43$  & $-0.80\pm 0.38$   &&    $ 0.84\pm 0.46$ &  $-0.64\pm 0.46$ &  $-0.39\pm 0.42$   \\
\hline
& \multicolumn{7}{c}{M42}\\
28.4 &      $ 1.93\pm 0.03$ &  $-0.22\pm 0.03$  & $-0.34\pm 0.02$   &&    $ 1.01\pm 0.02$ &  $-0.08\pm 0.02$ &  $-0.16\pm 0.02$  \\
44.1 &      $-0.06\pm 0.09$ &  $ 0.15\pm 0.09$  & $ 0.06\pm 0.08$   &&    $ 0.29\pm 0.08$ &  $ 0.01\pm 0.08$ &  $-0.07\pm 0.07$  \\
70.4 &      $-0.83\pm 0.14$ &  $ 0.38\pm 0.14$  & $ 0.36\pm 0.12$   &&    $-0.12\pm 0.13$ &  $ 0.69\pm 0.14$ &  $ 0.78\pm 0.10$  \\
\hline
 & \multicolumn{7}{c}{Cygnus\,X}\\
28.4 &      $-2.18\pm 0.04$ &  $ 0.05\pm 0.03$  & $ 0.19\pm 0.04$   &&    $-0.94\pm 0.06$ &  $ 0.22\pm 0.03$ &  $ 0.20\pm 0.03$   \\
44.1 &      $-0.04\pm 0.05$ &  $-0.17\pm 0.05$  & $-0.07\pm 0.04$   &&    $ 0.25\pm 0.05$ &  $ 0.52\pm 0.05$ &  $ 0.46\pm 0.05$   \\
70.4 &      $ 0.73\pm 0.06$ &  $-0.62\pm 0.06$  & $-0.02\pm 0.06$   &&    $ 0.05\pm 0.06$ &  $-0.04\pm 0.06$ &  $-0.47\pm 0.06$   \\
\hline
\end{tabular}
\end{center}
\label{tab:lfi_leakage}
\end{table*}

\end{document}

%% file: quijote_acknow.tex

We thank the staff of the Teide Observatory for invaluable assistance in the commissioning and operation of QUIJOTE.
The {\it QUIJOTE} experiment is being developed by the Instituto de Astrofisica de Canarias (IAC),
the Instituto de Fisica de Cantabria (IFCA), and the Universities of Cantabria, Manchester and Cambridge.
Partial financial support was provided by the Spanish Ministry of Science and Innovation 
under the projects AYA2007-68058-C03-01, AYA2007-68058-C03-02,
AYA2010-21766-C03-01, AYA2010-21766-C03-02, AYA2014-60438-P,
ESP2015-70646-C2-1-R, AYA2017-84185-P, ESP2017-83921-C2-1-R,
PGC2018-101814-B-I00, PID2019-110610RB-C21, PID2020-120514GB-I00, 
IACA13-3E-2336, IACA15-BE-3707, EQC2018-004918-P, the Severo Ochoa Programs SEV-2015-0548 and CEX2019-000920-S, the
Maria de Maeztu Program MDM-2017-0765, and by the Consolider-Ingenio project CSD2010-00064 (EPI: Exploring
the Physics of Inflation). We acknowledge support from the ACIISI, Consejeria de Economia, Conocimiento y 
Empleo del Gobierno de Canarias and the European Regional Development Fund (ERDF) under grant with reference ProID2020010108, and 
Red de Investigaci\'on RED2022-134715-T funded by MCIN/AEI/10.13039/501100011033.
This project has received funding from the European Union's Horizon 2020 research and innovation program under
grant agreement number 687312 (RADIOFOREGROUNDS), and the Horizon Europe research and innovation program under GA 101135036 (RadioForegroundsPlus).